\documentclass[english,twocolumn,prl,superscript]{revtex4-1}
\usepackage[T1]{fontenc}
\usepackage[latin9]{inputenc}
\usepackage{color}
\usepackage{array}
\usepackage{multirow}
\usepackage{amsmath}
\usepackage{amssymb}
\usepackage{graphicx}
\usepackage{wasysym}

\makeatletter

\providecommand{\tabularnewline}{\\}

\makeatother

\usepackage{babel}
\usepackage{listings}

\begin{document}
\title{Rapid Exploration of Topological Band Structures using Deep Learning}
\author{Vittorio Peano}
\affiliation{Max Planck Institute for the Science of Light, Erlangen, Germany}
\author{Florian Sapper}
\affiliation{Max Planck Institute for the Science of Light, Erlangen, Germany}
\affiliation{Department of Physics, Friedrich-Alexander Universität Erlangen-Nürnberg,
Germany}
\author{Florian Marquardt}
\affiliation{Max Planck Institute for the Science of Light, Erlangen, Germany}
\affiliation{Department of Physics, Friedrich-Alexander Universität Erlangen-Nürnberg,
Germany}
\begin{abstract}
The design of periodic nanostructures allows to tailor the transport
of photons, phonons, and matter waves for specific applications. Recent
years have seen a further expansion of this field by engineering topological
properties. However, what is missing currently are efficient ways
to rapidly explore and optimize band structures and to classify their
topological characteristics, for arbitrary unit cell geometries. In
this work, we show how deep learning can address this challenge. We
introduce an approach where a neural network first maps the geometry
to a tight-binding model. \textcolor{black}{The tight-binding model
encodes not only the band structure but also the symmetry properties
of the Bloch waves. This allows us to rapidly categorize a large set
of geometries in terms of their band representations, identifying
designs for fragile topologies. We demonstrate that our method is
also suitable to calculate strong topological invariants, even when
(like the Chern number) they are not symmetry-indicated.  Engineering
of domain walls and optimization are accelerated by orders of magnitude.
Our method directly applies to any passive linear material, irrespective
of the symmetry class and space group. It is general enough to be
extended to active and nonlinear metamaterials.}
\end{abstract}
\maketitle

\section{Introduction }

Wave propagation in a periodic medium is governed by a band structure
that substantially modifies the transport of those waves. While these
effects were first explored for electrons inside crystals, with the
atomic arrangement dictated by chemistry, band structures are also
encountered in many other areas across physics where modern advances
make it possible to engineer the periodic medium: photonic \citep{joannopoulos_photonic_2008}
and phononic \citep{maldovan_sound_2013} crystals as well as optical
lattices \citep{cooper_topological_2019} are well-known examples.
This offers the opportunity to explore freely the space of possible
designs and search for band structures with peculiar desired properties. 

One particularly exciting target for such explorations are the topological
features that have become a centerpiece of modern band structure theory
\citep{hasan_colloquium:_2010,ozawa_topological_2019,cooper_topological_2019,bradlyn_topological_2017}.
Recent theoretical breakthroughs \citep{po_symmetry-based_2017,bradlyn_topological_2017}
have allowed the exploration of large databases of natural materials
to uncover thousands of topological materials \citep{vergniory_complete_2019}.
For engineered materials, on the other hand, the configuration space
is even infinite-dimensional. There, an efficient method to rapidly
extract the band structure and topology for any given unit cell geometry
would be a crucial tool which could pave the way to discoveries that
would otherwise not be feasible. Ideally, such a method should (i)
provide answers for completely arbitrary geometries, (ii) be easily
transferrable to different underlying wave equations, (iii) allow
a substantial speed-up compared to state-of-the art methods, and (iv)
predict topological properties.

We believe that deep learning approaches are uniquely suited to address
these challenges.  Up to now, the first applications of neural networks
to band structures have focused on learning the mapping of a few selected
model parameters (describing the geometry of the periodic medium)
to the bands \citep{malheiros-silveira_prediction_2012,pilozzi_machine_2018,ferreira_computing_2018,shi_deep_2019},
band gaps \citep{zhaochun_artificial_1998,schmidt_recent_2019}, or
topological invariants \citep{zhang_machine_2018,sun_deep_2018,claussen_detection_2019}.
However, NNs can clearly be designed to make predictions for arbitrary
unit-cell geometries, enabling the exploration of a much wider design
space. This is closely related to the well-developed domains of image
recognition and image-to-image mapping. While that would already be
an important step on its own, such a NN would still be oblivious of
any property imprinted in the Bloch waves, including any topological
property. 

\begin{figure*}
\includegraphics[width=0.8\paperwidth]{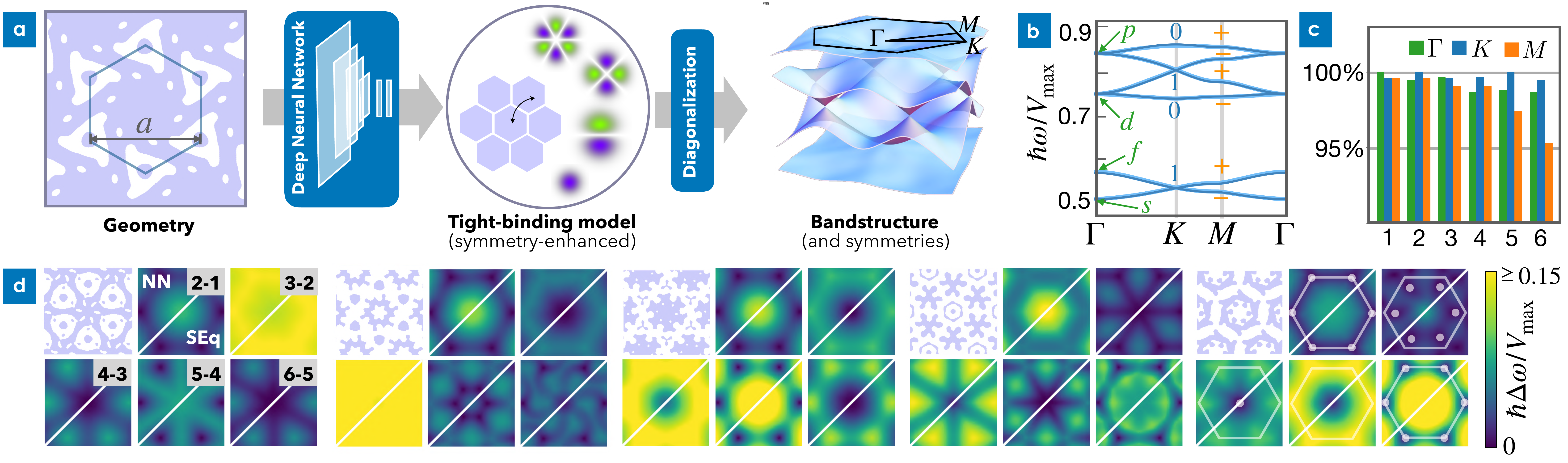}

\caption{\label{fig:Neural-network-based-prediction-}Neural-network-based
prediction of band structures from unit cell geometries or potentials.
(a) The geometry is fed into a multi-layer convolutional network (Appendix
\ref{sec:Network-layout}), producing the coefficients of a symmetry-enhanced
tight-binding model, which is then diagonalized to obtain the band
structure. (b) Cut along the $\mathbf{k}$-path indicated in (a),
for the same band structure (now including also the first two bands),
comparing Schrödinger equation (dark thin lines) and NN (thick light
lines). \textcolor{black}{The two types of predictions are difficult
to distinguish with the bare eye.}\textbf{ }The symmetry labels are
also indicated (0/1 indicate the quasi-angular momentum; $+/-$ label
even/odd states). (c) Fraction of correctly predicted symmetry labels,
at each high-symmetry point ($\Gamma$, $K$, and $M$), and for each
band, on validation geometries unknown to the NN. (d) Comparing the
$\mathbf{k}$-resolved band gap $\omega_{n+1}(\mathbf{k})-\omega_{n}(\mathbf{k})$
for five validation geometries (NN vs. Schrödinger). For the last
potential, the band gap zeros are marked by white dots. The crossings
of the second and third band define a sextuplet of Dirac cones that
are not pinned to any high-symmetry point. In all plots, the height
of the potential is $V_{{\rm max}}=(16\hbar)^{2}/(2ma^{2}).$}
\end{figure*}

The solution we advocate here is to have the neural network (NN) turn
an arbitrary unit cell geometry into the parameters of a tight-binding
(TB) model (see Fig.~\ref{fig:Neural-network-based-prediction-}).
In a subsequent step, this small TB model is then efficiently diagonalized
to yield the full band structure as well as the topologically relevant
features of the Bloch waves. Essential constraints imposed by the
symmetries of the underlying geometry can be straightforwardly implemented
in such a TB model. The whole approach is an example of 'known-operator-learning'
\citep{maier_learning_2019}, where one embeds into a NN a function
that implements a complex (but known) operation that is useful in
the given context.

We show that the rapid exploration made possible by our NN is a powerful
tool to aid in physical discovery. It addresses challenges in design
and optimization, answering questions like: Is it possible to implement,
under given physical constraints, a band structure of interest --
e.g. as produced by a simpler toy model? If yes, which combinations
of model parameters are accessible? \textcolor{black}{How abundant
are topological bands for an arbitrary distribution of designs? What
is the distribution of a topological invariant like the Chern number?}

\section{Scope and Case Studies}

\textcolor{black}{Our approach has a broad scope as it directly applies
to any linear metamaterial supporting band structures encoded in Hermitian
Hamiltonians. As such, it is applicable to electronic, photonic, phononic
systems and beyond. For each case study, the symmetry group can be
arbitrarily chosen but it remains fixed. In other words, in our method
a NN is trained to predict a distribution of band structures which
share one of the 230 space groups (or in 2D, one of the $17$ wallpaper
groups) and one of the $10$ symmetry classes. The latter accounts
for so-called generalized symmetries: the time-reversal, the particle-hole
and the chiral symmetries \citep{altland_nonstandard_1997}. }

\textcolor{black}{Below, we present the essential elements of our
method in a general framework. For the sake of concreteness and to
prove its practical value, we also discuss in detail a few interesting
case studies. The case study which we have used to demonstrate most
applications is the 2D Schrödinger equation $[-\hbar^{2}\triangle/2m+V(\mathbf{x})]\psi_{n}=\hbar\omega_{n}\psi_{n}$
with ${\cal C}_{6}$ symmetric (translationally invariant) potentials.
This corresponds to the $p6$ wallpaper group and the symmetry class
AI (conserved time-reversal symmetry whose square is the identity
and no particle-hole or chiral symmetry). Inspired by the situation
that is encountered in photonic or phononic crystals, where the geometry
of two materials (solid/air) defines the unit cell we have focused
on step-like potentials. This relates specifically to the propagation
of light in photonic-crystal type optical waveguides in the paraxial
approximation, which generally has been an important playground for
photonic topological physics recently \citep{kraus_topological_2012,rechtsman_photonic_2013,weiss_topological_2013,bandres_topological_2016,russell_helically_2017}.
We note that the symmetry class AI does not support any strong topological
invariant in 2D but still allows for fragile topological phases \citep{po_fragile_2018}.
Below we will show that our method is especially well suited to identify
this type of topological phases.}

\textcolor{black}{Furthermore, we demonstrate that its realm of applications
extends to systems supporting strong topological phases. For this
purpose, we have considered as a case study the $2$D Dirac equation
with a position-dependent mass, see Section \ref{sec:Strong-topological-phases}.
This equation has particle-hole symmetry (squaring to the identity)
but broken time-reversal symmetry and, thus, belongs to the symmetry
class D, which supports topological phases with non-trivial Chern
numbers. This model is of interest on its own as it captures the large-wavelength
spin-polarized (or mirror-symmetry-polarized) physics of a HgTe/CdTe
quantum well \citep{konig_quantum_2007} whose geometrical parameters
are varied periodically to realize a pattern of alternating trivial
and topological insulator domains.}

\textcolor{black}{Finally, we demonstrate another aspect of the flexibility
of our method, implementing it for the 3D Schrödinger equation in
the presence of potentials with a non-symmorphic space group (the
space group $p4_{2}22$).}

\textcolor{black}{In the context of applications of deep neural networks
for topology, our band-structure-based approach, with direct predictions
based on the underlying geometry, is of a different nature than other
approaches where the network tries to identify (topological) phases
of matter based on observing e.g. simulated snapshots of system configurations
or correlators \citep{carrasquilla_machine_2017,deng_machine_2017,zhang_quantum_2017,lian_machine_2019,schafer_vector_2019}.}

\section{Tight-binding Neural Network }

In the standard setting of band structure theory, a wave equation
is solved on a periodic lattice, giving rise to a set of bands $\omega_{n}(\mathbf{k})$,
where $n$ is the band index and $\mathbf{k}\in{\rm BZ}$ the wave
vector inside the Brillouin zone. The waves are subject to a periodic
modulation of a potential (in the case of the Schrödinger equation),
a dielectric index (for the Maxwell equations), or material density
and elastic moduli (for phononic crystals). To keep our description
general, we will simply refer to 'the unit cell geometry' in either
case.

In our case, we propose to use the NN, to generate a TB Hamiltonian:
$\hat{H}=\hat{H}(F_{\theta}(V(\cdot)))$. Here $V(\cdot)$ represents
the network's input (the unit cell geometry; i.e. a potential or a
material distribution), $\theta$ is a vector collecting all the network's
parameters (weights and biases), and $F_{\theta}$ is the network's
output: a vector that contains the energies and hopping matrix elements
of the TB model. 

The band structure, in turn, results from writing this Hamiltonian
in $k$-space, and diagonalizing the resulting $N\times N$ matrix
$\hat{H}_{\mathbf{k}}(F_{\theta}(V(\cdot)))=\left\langle \mathbf{k}\left|\hat{H}(F_{\theta}(V(\cdot)))\right|\mathbf{k}\right\rangle $.
The number $N$ of TB orbitals is chosen depending on how many bands
we would like to predict, and we will comment more on this later.
Overall, for any given wave vector $\mathbf{k}$, we generate a vector
$\mathbf{\omega}=(\omega_{1},\omega_{2,}\ldots,\omega_{N})$ of eigenfrequencies,

\[
\hbar\mathbf{\omega}(\mathbf{k})={\rm \mathbf{Diag}}[\hat{H}_{\mathbf{k}}(F_{\theta}(V(\cdot)))]
\]
As indicated above, it is important for network training that the
diagonalization operation ${\rm \mathbf{Diag}}$ is differentiable
with respect to the entries of the Hamiltonian matrix. Indeed, from
first-order Rayleigh-Schrödinger perturbation theory one finds

\begin{equation}
\frac{\partial\omega_{n}(\mathbf{k})}{\partial\theta}=\sum_{l}\left\langle \phi_{n}(\mathbf{k})\left|\frac{\partial\hat{H}_{\mathbf{k}}(F_{\theta}(x))}{\partial F_{\theta}^{(l)}}\right|\phi_{n}(\mathbf{k})\right\rangle \frac{\partial F_{\theta}^{(l)}(x)}{\partial\theta}\,,\label{eq:DifferentiateDiag}
\end{equation}
Here $\left|\phi_{n}(\mathbf{k})\right\rangle $ is the eigenvector
in the basis of TB orbitals, $\hat{H}_{\mathbf{k}}\left|\phi_{n}(\mathbf{k})\right\rangle =\omega_{n}(\mathbf{k})\left|\phi_{n}(\mathbf{k})\right\rangle $,
and $F_{\theta}^{(l)}$ are the parameters inside the tight-binding
Hamiltonian that have been predicted by the NN.

The cost function during training is prescribed as the quadratic deviation
between the true band structure and the predictions obtained from
the network, averaged over all training samples $V(\cdot)$, the bands
$n,$ and the quasimomentum $\mathbf{k}$:

\begin{equation}
C(\theta)=\left\langle \left(\omega_{n}^{{\rm NN}}(\mathbf{k})-\omega_{n}^{{\rm true}}(\mathbf{k})\right)^{2}\right\rangle _{V(\cdot),n,\mathbf{k}}\label{eq:cost}
\end{equation}
The set of $\mathbf{k}$-points is a grid covering the full Brillouin
zone (BZ). See Appendix \ref{sec:Gradient-descent-for} for details
on the implementation of the resulting gradient descent (using TensorFlow).

\subsection{Symmetry-enhanced tight-binding model. }

One of the important advantages of this approach is the ability to
take care of the space group and other symmetries in an elegant and
efficient way, by imposing them on the TB model. \textcolor{black}{This
is particularly important for topological band structures whose topological
features are well known to be constrained (and in some cases even
determined) by the underlying symmetry properties.}

\textcolor{black}{We will call such a TB model ``symmetry-enhanced''.
This TB model shares the same space group and symmetry class as the
training samples. In order to define its Hilbert space, we select
a basis of localized Wannier orbitals. The choice of a suitable set
of orbitals depends not only on the space group and symmetry class
but also on the potential distribution and the number of bands we
would like to predict.} The space group and the generalized symmetries
impose constraints on the hopping and onsite energies of our TB model
(that depends on localization position and point symmetry of the orbitals).
Each output neuron of our NN encodes an independent parameter of the
underlying Hamiltonian $\hat{H}$, \textcolor{black}{see Appendix
\ref{sec:Symmetry-enhanced-tight-binding} for more details}\textcolor{red}{. }

During training, we require that the Bloch wave symmetries at a discrete
set of so-called maximal ${\bf k}$-points (e.g. $\Gamma$, $K$ and
$M$ for the $p6$ group) are reproduced correctly. This also ensures
the correct behaviour at all other high-symmetry points or lines \citep{bradlyn_topological_2017}
that may occur in general for arbitrary space groups. For these ${\bf k}$-points,
the Hamiltonian decomposes into blocks corresponding to an irreducible
representation (irrep) of the proper symmetry group of ${\bf k}$
\textcolor{black}{(for non-symmorphic groups the little group, see
Section \ref{sec:Application-to-a}).}\textcolor{red}{{} }In practice,
we enforce the right behavior by applying the cost function (\ref{eq:cost})
separately to each block at the maximal $\mathbf{k}$-points -- demanding
a match to the training data for each symmetry sector separately. 

\section{Training }

\label{sec:Training}An important challenge in NN training is the
choice of training data. If data are generated by simulation (as is
the case here), one can train on random input with a distribution
close to the envisaged applications. Our approach has been to generate
Gaussian random fields in the unit cell (with independent Fourier
components, here $\left\langle \left|A_{\mathbf{k}}\right|^{2}\right\rangle \sim\left|\mathbf{k}\right|^{-1}$,
see Appendix \ref{sec:Generating-Training-Samples}). The Fourier
components are enforced to be of the appropriate symmetry. When required,
step-like potentials can be implemented by digitizing the initially
continuous random field to two values, $V(\mathbf{x})=0$ or $V(\mathbf{x})=V_{{\rm max}}$. 

For the 2D Schrödinger equation case study, we have trained the network
on the six lowest energy bands, using a $\mathbf{k}$-grid inside
a triangular region covering uniformly $1/6$th of the Brillouin zone
(sufficient for ${\cal C}_{6}$ symmetry), see Appendix \ref{sec:Training-of-the}
for more details. As discussed above, afterwards our symmetry-enhanced
TB model still allows us to predict the band structure with arbitrary
$\mathbf{k}$-space resolution. The results produced using our NN
(Fig.~\ref{fig:Neural-network-based-prediction-}) are essentially
indistinguishable from the true bands: our NN can predict the band
structure with about $2\%$ accuracy (relative to typical band gaps
Appendix \ref{sec:Accuracy-for-the}), after training on 50,000 samples,
and it is about 1000 times faster than Lanczos-type diagonalization.
For a more detailed discussion of the performance gain allowed by
our NN, see Appendix \ref{sec:Performance-Gain}.

In addition, our NN also predicts \textcolor{black}{the underlying
proper-group irreps for the Bloch waves at the maximal $\mathbf{k}$-points.
Throughout the paper, we refer to the labels identifying such irreps
as symmetry labels.} We take the example of the $p6$ group to illustrate
how symmetries automatically give rise to robust features of the band
structure that would be difficult to predict otherwise. For $p6$,
the proper group for each maximal ${\bf k}$-point is a rotational
group ${\cal C}_{n}$, with $n=6$ for $\Gamma$, $n=3$ for $K$,
and $n=2$ for $M$, cf. Fig.~\ref{fig:Neural-network-based-prediction-}.
The combination of time-reversal symmetry and ${\cal C}_{n}$ rotations
gives rise to robust features: (i) At the $\Gamma$-point, $p$ and
$d$ Bloch waves come in pairs with opposite quasi-angular momentum
and lead to parabolic band touching (Fig.~\ref{fig:Neural-network-based-prediction-}b).
(ii) Likewise, at the $K$ points, essential degeneracies arise from
pairs of states with opposite quasi-angular momentum $m_{K}=\pm1$,
leading to Dirac cones, cf Fig. \ref{fig:Neural-network-based-prediction-}(b,d). 

We emphasize that such features are automatically enforced by our
symmetry-enhanced TB model. This is one of its main advantages over
a naive approach. A statistical analysis of a set of validation samples
shows that the fraction of correctly predicted symmetry labels is
about $99\%$, cf Fig.~\ref{fig:Neural-network-based-prediction-}(c).
\textcolor{black}{We have checked that this is only limited by the
rms band structure deviation, see also Section \ref{sec:Application-to-a}.
In other words, the NN exchanges the ordering of two levels with different
symmetry labels only if their splitting happens to be so small that
the NN is not able resolve it. }

The central focus of modern band structure theory is the study of
topological properties. These cannot be deduced from the band structure
$\omega_{n}(\mathbf{k})$ itself, but only from the behaviour of Bloch
waves. We will show that, remarkably, our NN learns to predict correctly
such properties despite having only very limited implicit information
regarding the Bloch waves (via the symmetries). This is crucial, because
training for the full eigenstates throughout the BZ would drastically
increase the size of the NN and slow down training.

\begin{figure}
\includegraphics[width=1\columnwidth]{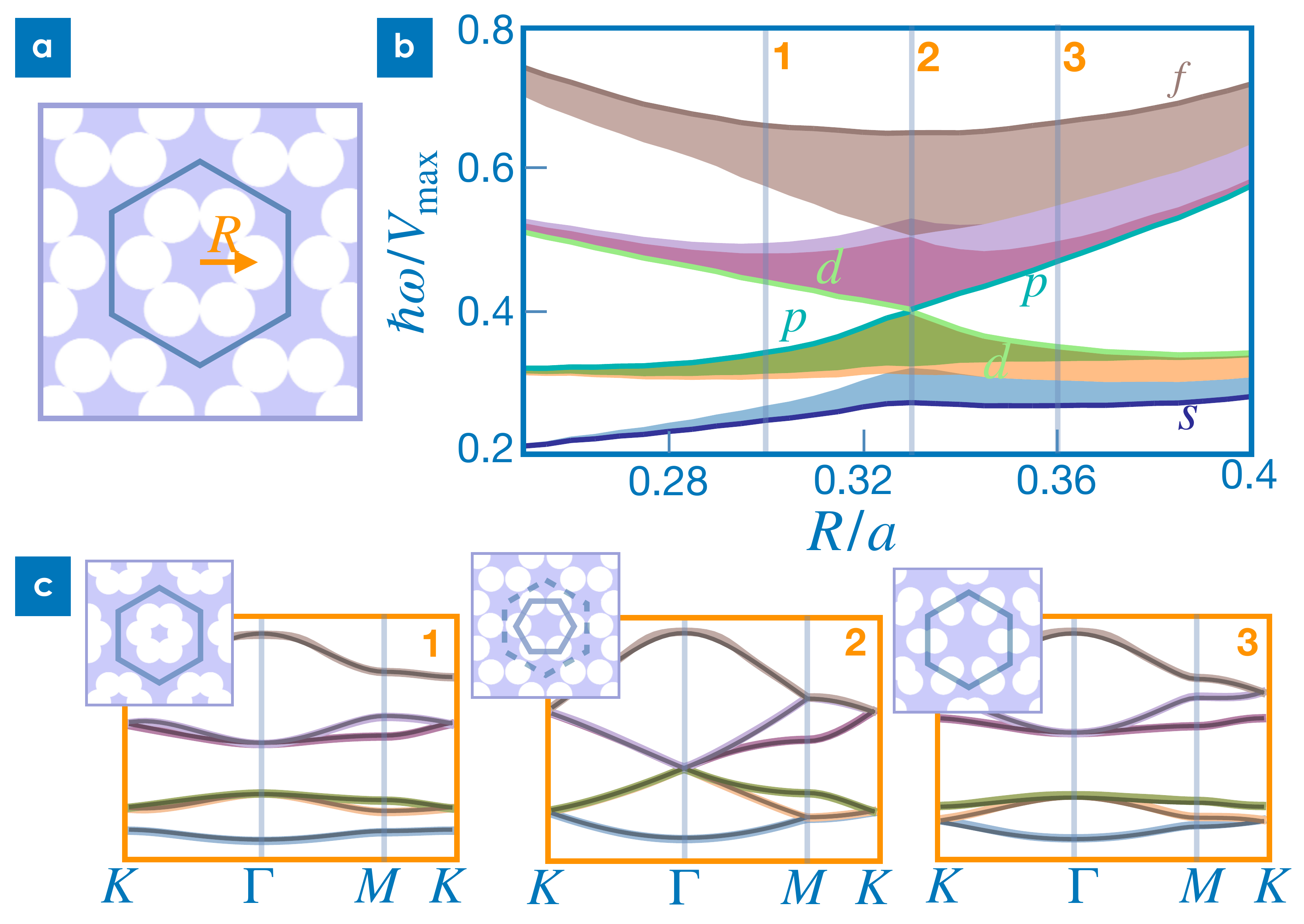}

\caption{\label{fig:WuandHu}Designing a band inversion for topological transport,
using rapid band structure evaluation and symmetry predictions provided
by a neural network. (a) Geometry of the potential: six circular
holes of fixed radius\textbf{} are placed at a varying distance $R$
from the ${\cal C}_{6}$ centre. (b) Energy spectrum at the $\Gamma$-point
as a function of $R$. The energies of the $p$ and $d$ bands cross
for $R=a/3$ (c) Band structure for three different values of $R$
(marked in panel (b) by the horizontal lines) before, at, and after
the band inversion transition (thick lines: NN, thin lines: SEq).
The corresponding potentials are shown as insets. At the band inversion
transition the Wigner-Seitz primitive cell becomes smaller. The resulting
folded band structure supports a pair of degenerate Dirac cones at
the $\Gamma$-point. Moreover, two pairs of bands become degenerate
along the $\mathbf{k}$-path from $M$ to $K$. These are essential
degeneracies enforced by the rotational symmetry and the smaller unit
cell. The NN is able to reproduce these features although it has not
been trained on potentials with a smaller unit cell.}
\end{figure}

\section{Design of band inversions }

The bulk-boundary correspondence provides a link from the bulk topology
to the existence of robust gapless excitations at a physical boundary
or domain wall. This paves the way to using a NN that has been trained
on the bulk band structure and Bloch wave symmetries as a tool to
design topological edge states.

For topological insulators, a generic mechanism leading to a non-trivial
topology and helical edge states is the so-called band inversion in
which the usual ordering of a pair of bands is exchanged. For photonic
and phononic crystals, a band inversion of $p$ and $d$ orbitals
can be engineered by purely geometrical means \citep{wu_scheme_2015}.
Based on this concept, our NN helps to efficiently design domain walls
of this type. In Fig.~\ref{fig:WuandHu} the geometry is tuned to
decrease the energy of a $d$-orbital while increasing the energy
of a $p$-orbital until their order is inverted. The very close agreement
between the network predictions and the true spectrum is remarkable,
given that the potential designs adopted here look very different
from the random training potentials. 

\begin{figure*}
\includegraphics[width=0.8\paperwidth]{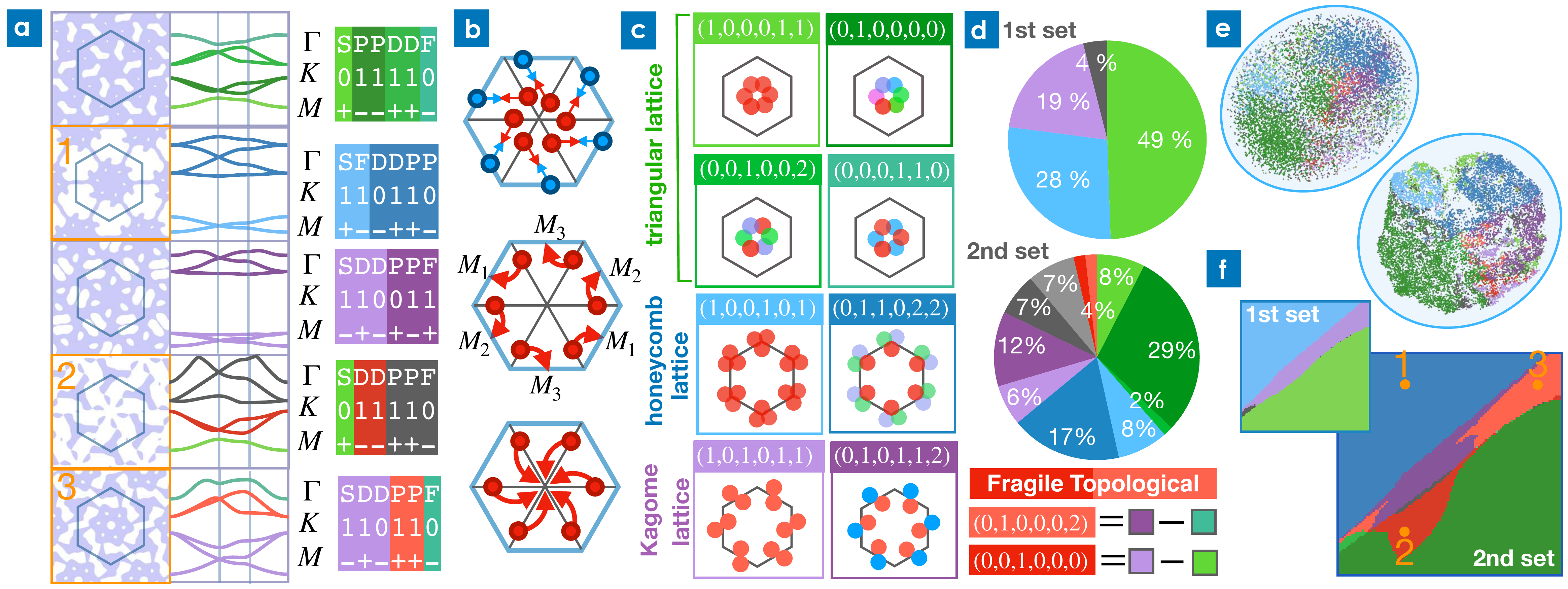}

\caption{\label{fig:Topology}Using the NN to explore the topological properties
of large sets of potentials. (a) Examples of randomly generated potentials,
and the band structure predicted by the NN. The tables indicate the
irreps predicted by the NN for each of the first 6 bands. (b) Possible
pathway leading to the annihilation of two (one) sextuplets of Dirac
cones. For one sextuplet this is only possible at $\Gamma$ or $M$.
(c) The EBRs for the wallpaper group $p_{6}$. The sketches depict
the 'Wyckoff' positions and point symmetry of the underlying orbitals.
The 'occupations' $n_{i}$ in the 'symmetry fingerprint' $(n_{s},n_{p},n_{d},n_{f},n_{0},n_{+})$
count the number of (pairs of) Bloch waves for the corresponding irreps,
e.g. $n_{p}=1$ means that there is one pair of $p$-waves at $\Gamma$.
(d) Distribution of (quasi-)BRs for the first (upper chart) and second
(lower chart) set of bands. The light grey slice represents sets of
bands that cannot be classified based on the first $6$ bands only
but are likely to be composite BRs, see Appendix \ref{sec:AppTopexpl}.
All other composite BRs are grouped in the dark grey slice. (e) t-SNE
visualization, with each point representing a random potential (left)
or the corresponding output (TB coefficients) of the NN (right); the
latter allows for improved clustering (colors indicate EBRs). (f)
Phase map indicating EBR and fragile topology, for the first (inset)
and the second sets of bands. The parameter space interpolates between
the potentials '$1$', '$2$', and '$3$' marked in panel (a), in
terms of their underlying Fourier coefficients. }
\end{figure*}

\section{Exploring Band Representations and Fragile Topological phases }

Topological band structure theory originally relied entirely on momentum-space
properties, defining topological invariants based on the behaviour
of Bloch waves across the Brillouin zone. Only relatively recently,
it was realized that important additional information can be extracted
by analyzing the tension between momentum-space and real-space descriptions.
The resulting mathematical theories \citep{po_symmetry-based_2017,bradlyn_topological_2017,cano_building_2018,vergniory_complete_2019}
(sometimes known as ``topological quantum chemistry'') build onto
the theory of band representations \citep{zak_symmetry_1980} to offer
a very general theoretical framework to classify all natural materials
according to their topological properties. This theoretical formalism
has been so far mostly used to investigate electronic properties of
natural materials. However, its range of potential applications extends
to any periodic medium, see Ref. \citep{de_paz_engineering_2019}
for a pioneering application to photonics. Here, we demonstrate how
our NN based approach combined with topological quantum chemistry
allows the rapid exploration and statistical analysis of the topological
properties of large sets of band structures. 

Band representation (BR) theory tries to understand isolated sets
of bands (separated from the remaining bands everywhere by local gaps)
in terms of their underlying Wannier orbitals. Mathematically, a BR
is a (time-reversal symmetric) space group representation that is
defined on a basis of Wannier states in the so-called atomic limit
\citep{cano_building_2018}. Intuitively, this corresponds to the
limit where all Wannier states have a localization length that is
much shorter than the lattice length scale. A group of bands corresponds
to a BR if it is possible to reach the atomic limit by continuously
modifying the Hamiltonian without closing the relevant band gaps.
For topological bands, it is not possible to reach the atomic limit
under continuous deformations. Topological quantum chemistry aims
to identify materials hosting such bands. Remarkably, in most cases
this is possible based solely on the band structure and the irreps
at the maximal $\mathbf{k}$-points, exploiting the fact that all
BRs can be decomposed in terms of building blocks known as elementary
band representations (EBRs) \citep{zak_symmetry_1980}. Crucially
this information is also made available by our NN (Fig.~\ref{fig:Topology}a). 

We demonstrate the power of the NN by analyzing randomly generated
potentials. They are sampled from a distribution which, in practical
applications, might be dictated by experimental design constraints.
Here, we illustrate it for $\left\langle \left|A_{\mathbf{k}}\right|^{2}\right\rangle \sim\left|\mathbf{k}\right|^{-1/2}$.
Even though this is different from the training distribution, the
network performs very well. 

In a first step, one needs to identify isolated sets of connected
bands, which in topological quantum chemistry is commonly done by
checking for connections only at high-symmetry points. Our approach
allows to go beyond that by efficiently searching for connections
\emph{away} from these points -- looking for $\pi$-defects in the
Berry flux on a fine ${\bf k}$-grid (much finer than the training
grid), evaluated rapidly thanks to the small Hilbert space of the
NN-generated tight-binding model. In this way, we can easily scan
large ($\sim10^{4}$) sets of potentials using this method that would
be otherwise computationally expensive. Our numerical results show
that any clustering of bands based only on connections at high-symmetry
points would be incorrect for a substantial fraction of the potentials
($\sim10\%$ for the second set of connected bands). In most cases,
this error translates into a wrong topological classification of the
bands, see below. 

Inspired by these observations, we set out ourselves to investigate
how robust are the connections away from high-symmetry points. More
precisely, we wondered whether -- as is often assumed, e.g. \citep{po_symmetry-based_2017}
-- it is possible to eliminate them without re-arranging the order
of bands at those points. Band touchings are protected by the ${\cal C}_{2}{\cal T}$
anti-unitary symmetry \citep{van_miert_dirac_2016} and can, thus,
be eliminated only by pairwise cone annihilation. This led us to distinguish
two scenarios: (i) If an \emph{odd} number of cones is present in
$1/6$th of the BZ, the cones can be annihilated only at the $\Gamma$-point
or at the $M$-points, cf Fig.~\ref{fig:Topology}c. This implies
a re-arrangement of the band order at the high-symmetry points. (ii)
Otherwise (for an even number) the cones can be annihilated anywhere
(Fig.~\ref{fig:Topology}c), without re-arrangement. Using the NN
we have discovered that the first scenario occurs in the overwhelming
majority of cases ($\approx95\%$, for our potential distribution).
\textcolor{black}{The presence of robust connections in this scenario
seems to point to a missing compatibility relation. Indeed, such a
relation can be identified as a consequence of a previous finding
in the literature \citep{fang_bulk_2012}. In our time-reversal invariant
system, the sum of the Chern numbers for a set of connected bands
is always zero. As shown in \citep{fang_bulk_2012}, the overall parity
of the ${\cal C}_{2}$ eigenvalues at the ${\cal C}_{2}$-symmetric
$\mathbf{k}$-points (the parity of the number of odd states for a
set of connected bands) is equal to to the parity of the Chern number,
which therefore means that every connected set of bands must have
overall even parity in our system.}

As a final step towards identifying topological sets of bands, we
enumerate all EBRs, assigning to each a unique symmetry fingerprint
($\mathbb{N}^{n}$ array) that lists the number of (degenerate) orbitals
for each irrep at each symmetry point (Fig.~\ref{fig:Topology}d)
\citep{kruthoff_topological_2017,po_symmetry-based_2017}. For the
group $p6$, the $8$ possible irreps at the $\Gamma,K,M$ points
result in $n=6$ -- by noting the constraints imposed by the appropriate
compatibility relations, see Appendix \ref{sec:Symmetry-Fingerprints}.
If the fingerprint computed for an isolated set of bands cannot be
written as a sum of such EBR-fingerprints, the set must be topological
(sometimes labeled ``quasi-BR'').

We have used our NN to determine (quasi)-BRs for $10^{4}$ potentials
(Fig.~\ref{fig:Topology}e). For $4\%$ of the samples in this distribution,
the second set of bands is topological. Strictly speaking, this figure
depends on the statistical distribution of potentials, but we expect
qualitatively similar behavior for other distributions, see Appendix
\ref{sec:AppTopexpl}. The standard analysis without taking into account
connections away from the high-symmetry points would overestimate
this figure significantly, predicting $14\%$ of topological samples.
On the other hand, it turns out that this discrepancy is eliminated
once the connections predicted by our ${\cal C}_{2}$ compatibility
relation are taken into account. In this case, one recovers with high
statistical precision the results already obtained using the much
more numerically expensive Berry flux method. This gives also a way
to check our results solving directly the Schrödinger equation, see
Appendix \ref{sec:AppTopexpl} for more details. Besides providing
statistical insights, our study also represents an efficient random
search, uncovering hundreds of topological samples. Moreover, we obtain
important qualitative information: all quasi-BRs discovered here belong
to one of two cases (\ref{fig:Topology}e), where the set of bands
is obtained by splitting a BR into a topological band and another
BR. This is the defining feature of the recently discovered fragile
topological phases \citep{po_fragile_2018,po_faithful_2019,de_paz_engineering_2019,peri_experimental_2020}. 

A further task rendered feasible by the NN is the creation of high-resolution
multi-dimensional maps that explore the topological and hybridization
phase transitions encountered while interpolating between potentials
(Fig.~\ref{fig:Topology}f). 

\begin{figure}
\includegraphics[width=1\columnwidth]{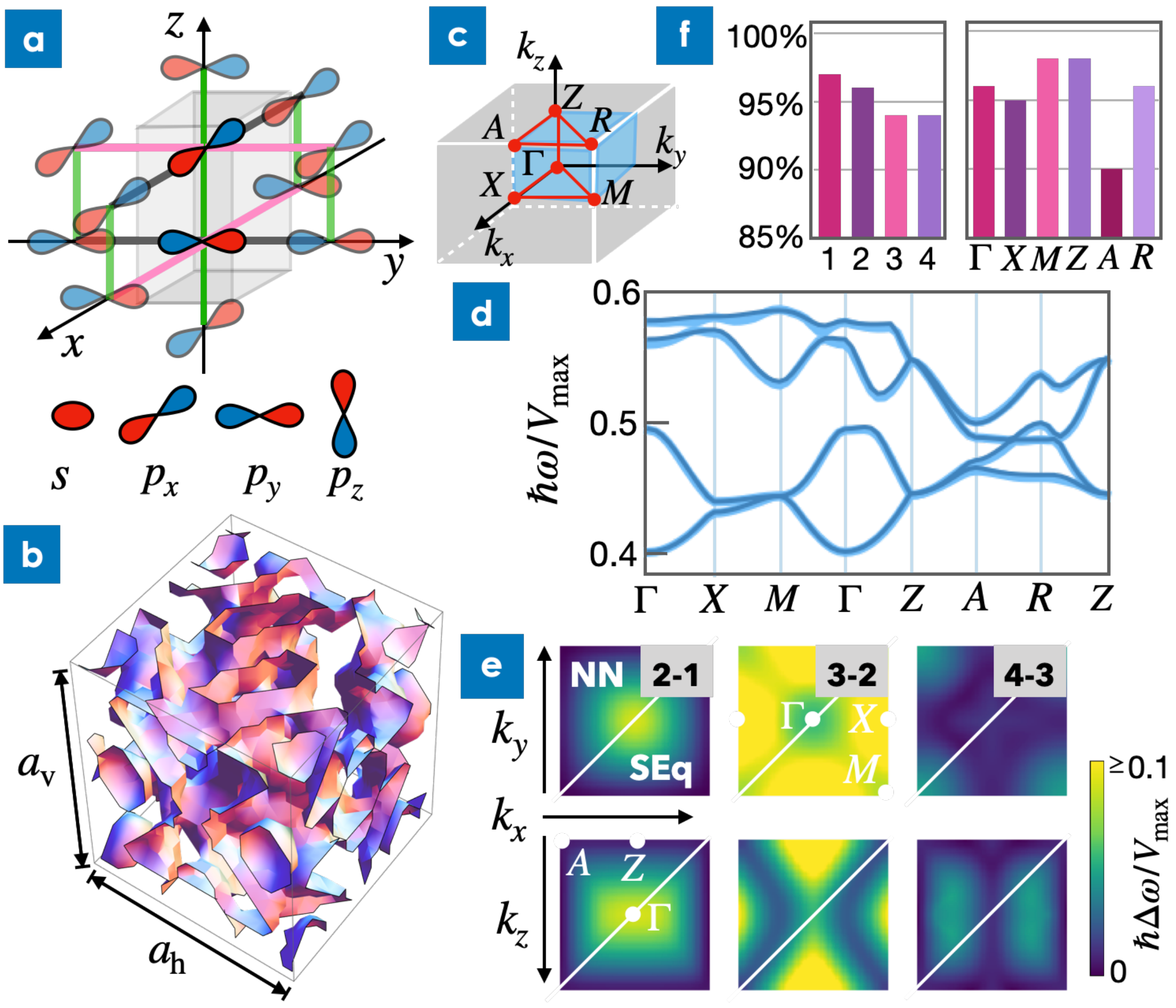}

\caption{\label{fig:Fig-3D} \textcolor{black}{Neural-network-based prediction
of 3D band structures for potential distributions in the $p4_{2}22$
non-symmorphic space group. (a) Sketch of the symmetry-enhanced TB
model. Pairs of orbitals localized about different sublattices are
mapped onto each other via screw rotations. Symmetry-related nearest-neighbors
transitions are highlighted in the same colour. The different orbital
types (bottom) transform according to the different representations
of the point group ${\cal D}_{2}$ (two-fold rotations about the axes
$x$, $y$, and $z$), see Appendix. (b) Example of step-like potential
with $p4_{2}22$ space group. (c) BZ and high-symmetry path for the
primitive tetragonal lattice. (d) Band structure along the high-symmetry
path. (e) Comparison of NN and Schrödinger equation predictions for
the first three band gaps on two different 2D cuts of the BZ. {[}The
results in (d) and (e) refer to the validation potential shown in
(b).{]} (f) Fraction of correctly predicted symmetry labels ordered
by band and high-symmetry point.}}
\end{figure}

\section{Application to a non-symmorphic example in 3D\label{sec:Application-to-a}}

\textcolor{black}{In this section we aim to demonstrate the flexibility
of our method by applying it to $3$D band structures. In doing so,
we switch our focus from wallpaper groups to space groups. Space groups,
unlike wallpaper groups, can not always be decomposed into a direct
sum of lattice translations and the point group. When this is not
possible, the space group is said to be non-symmorphic. For non-symmorphic
groups, the irreps classification -- an important step of our method
-- requires to take into account transformations that combine point
symmetries with translations by a fraction of a lattice vector, i.e.
screw rotations and glide mirrors. This is done generalizing the concept
of proper group by introducing the so-called little group. This is
an infinite dimensional subgroup of the space group that leaves invariant
a particular quasi-momentum and, in contrast to the proper group,
can include also some translations. It is also well known that TB
models with non-symmorphic space groups can only be implemented on
lattices with a basis \citep{landau_statistical_1980}. Thus, a lattice
with a basis will be required for our symmetry-enhanced TB model,
cf Fig \ref{fig:Fig-3D}(a). We remark that the majority of space
groups in $3$D are non-symmorphic (157 out of 230) and that non-symmorphic
space groups can even arise for quasi-2D systems, see e.g. \citep{young_dirac_2015}.
Motivated by the added complexity and the ubiquity of non-symmorphic
space groups we have decided to demonstrate our method for a non-symmorphic
example.}

\textcolor{black}{We consider the $3$D Schrödinger equation for step-like
potentials with space group $p4_{2}22$ (which has point group ${\cal D}_{4}$
and a primitive-tetragonal lattice), cf Fig \ref{fig:Fig-3D}(b) for
an example of such a potential. In addition to the lattice translations,
this group is generated by a screw rotation about the $z$-axis (a
rotation by $\pi/2$ accompanied by a translation by half a lattice
vector), and a two-fold rotation about the $x$-axis, cf Fig \ref{fig:Fig-3D}(a-b).
As usual we train our NN using a coarse grid, here, covering 1/8th
of the BZ, cf blue region in Fig \ref{fig:Fig-3D}(c). As explained
in Section \ref{sec:Symmetry-enhanced-tight-binding} and Appendix
\ref{sec:Details-of-the-3D}, the training cost function includes
a contribution for each of the irreps of the little groups at the
maximal $\mathbf{k}$-points. Here, we have six maximal $\mathbf{k}$-points
($\Gamma$, $M$, $X$ $Z$, $A$, and $R$) and a total of $28$
different irreps (substantially larger than our previous $p6$ example).
After taking the step-like ansatz for the potentials and rescaling
the energy, we are left with only two free parameters: the potential
height $V_{{\rm Max}}$ and the vertical lattice constant $a_{{\rm v}}$
(in appropriate units set by the horizontal lattice constant $a_{{\rm h}}$).
We have trained our NN to predict band structures for $V_{{\rm max}}=(10\hbar)^{2}/(ma_{{\rm h}}^{2})$
and $a_{{\rm v}}=a_{{\rm h}}$. In addition, we have discretized the
unit cell as a $20\times20\times20$ grid (comparatively low-resolution,
with the goal of saving computational resources in producing the training
data for this illustrative example).}

\textcolor{black}{The results are summarized in Fig. \ref{fig:Fig-3D}(d-f).
The band structures for validation potentials evaluated along a high-symmetry
path or high-resolution 2D cuts are in good agreement with exact results,
cf Fig. \ref{fig:Fig-3D}(d-e). More quantitatively, the rms deviation
of the band structure prediction averaged over $2000$ validation
samples is $2\times10^{-3}V_{{\rm Max}}$. The fraction of correctly
predicted symmetry labels is above $95\%$. The error decreases to
levels of around $1\permil$ if one takes into account only the levels
and symmetry-protected doublets that are separated by the neighboring
levels by more than three times the rms deviation. This indicates
that the fraction of correctly predicted symmetry labels is limited
only by the precision of the band structure predictions.}

\textcolor{black}{Overall, these results show that our method represents
a powerful tool to predict $3$D band structures that extends to systems
with very complex symmetry constraints including non-symmorphic systems.}

\section{Strong topological phases of two-component topological metamaterials
\label{sec:Strong-topological-phases}}

\textcolor{black}{In our approach to the prediction of band structures
and topological properties, the symmetry labels are the only information
about the underlying Bloch waves provided to the NN during training.
This approach has the advantage of being very efficient and it is
clearly suitable to predict symmetry-indicated topological features.
These also cover some (but not all) strong topological invariants.
It remains an open question whether our method can be adopted to investigate
non-symmetry-indicated topological features. The aim of this section
is to address this question. In particular, we focus on the most prominent
and well-known example of a non-symmetry-indicated strong topological
invariant, the Chern number for 2D systems.}

\textcolor{black}{A straightforward approach to predict Chern numbers
(or any other non-symmetry-indicated topological invariant) using
a NN consists in providing them to the NN during training and one-hot-encode
the NN's predictions in ad-hoc output neurons \citep{zhang_machine_2018,sun_deep_2018,claussen_detection_2019}.
While it would be possible to extend our method based on this approach,
we will instead use our standard method. However, this time, we will
train many NNs, on the same training data but with different random
initial conditions for the NN. We will then use the information regarding
the Chern numbers to postselect a NN that performs well on the Chern
numbers. This approach is far more elegant, as the Chern numbers are
then directly encoded in the symmetry-enhanced TB model. }

\textcolor{black}{At first sight, one may think that our approach
is doomed to fail: Since we aim to predict the Chern numbers for an
infinite number of geometries, one might naively expect that the number
of NNs that would be required to be trained should also be infinite.
However, one should keep in mind that the predictions of Chern numbers
for different geometries are highly correlated: Since a Chern number
can only change whenever the corresponding band gap closes, a NN that
is trained to accurately predict (via the symmetry-enhanced TB model)
the band structure will also automatically find the right topological
phase transition hypersurfaces in the space of all geometries. Hence,
if the NN is successfully postselected to predict the Chern numbers
for even a single geometry for each topological phase, it will automatically
provide the right Chern numbers for all other geometries as well.
In practice, the Chern numbers might still be wrong in those cases
where the band gaps are smaller than the precision of the NN. Only
the Chern numbers corresponding to band gaps much larger than the
precision will be guaranteed to be correct. However, since, as we
have seen, the NN precision is usually very high, this does not present
a siginificant problem in practice. The remainder of this section
is devoted to demonstrating that our expectations are fully confirmed,
for an interesting case study.}

\textcolor{black}{}
\begin{figure}
\textcolor{black}{\includegraphics[width=1\columnwidth]{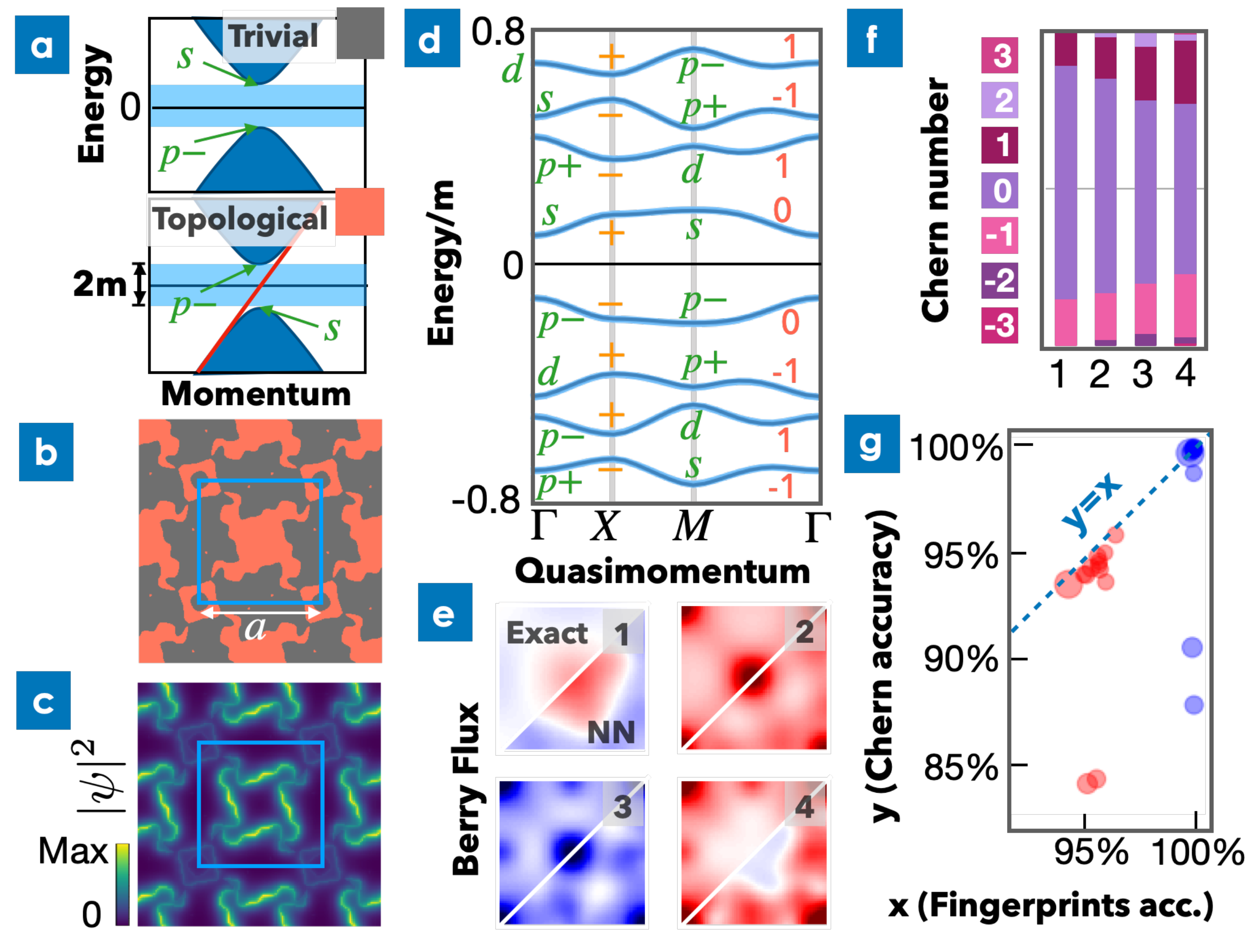}}

\textcolor{black}{\caption{\textcolor{black}{\label{fig:Chern} Using the NN to investigate the
spin Chern numbers in a family of 2D topological insulator metamaterials.
(a) Sketch of the spin-polarized excitation spectrum in a semi-infinite-plane
geometry for a trivial (top), and a topological (bottom) material
close to the $\Gamma$-point, close to the Fermi energy $E_{F}=0$.
The topological material supports gapless edge states (red line).
(b) Sketch of a metamaterial geometry. Its unit cell (blue) is subdivided
into trivial (grey) and topological (red) domains according to a randomly
drawn fourfold rotationally symmetric pattern. (c) Probability field
for a metamaterial's Bloch-wave with energy within the single-domain
band gap. (d) Metamaterial band-structure. Shown are the first (last)
four positive (negative) bands together with the symmetry labels and
the Chern numbers (in red). Exact results and NN predictions are not
distinguishable with bare eyes. (e) Exact and NN-predicted Berry fluxes
for the first four bands. {[}(c-e) refers to the validation geometry
shown in (b).{]} (f) Stacked bar chart of the Chern number distribution
for the first four positive bands. (g) Scatter plot of the Chern and
fingerprint accuracies for $13$ trained NNs. Each NN is represented
by a red and a blue circle. For the red circles the rms band structure
deviation (radius of the circles), the fraction of correct fingerprints
($x$-coordinate) and Chern numbers ($y$-coordinate) are obtained,
averaging over $2500$ validation samples and the eight central bands.
The blue circles take into account only bands separated from neighboring
bands by a minimal splitting larger than $0.01m$.}}
}
\end{figure}

\textcolor{black}{Even after restricting ourselves to 2D systems,
there is ample choice of possible case studies where the topological
phases are described by non-trivial Chern numbers. Formally, the only
precondition is that the relevant Hamiltonian has broken time-reversal-symmetry.
This is for example the case for charged particles in the presence
of a magnetic field or for certain systems under a suitable time-dependent
drive. Since the spin-orbit interaction is akin to a magnetic field
when acting on spin-polarized electrons, the Chern numbers are also
suitable to describe the topological phases of this type of excitations.
In this framework, the Hamiltonian of interest is 'half' (one of each
spin-polarized block) of the full time-reversal-symmetric spin-conserving
Hamiltonian. Thus, the non-trivial Chern numbers refer to only one
spin sector and are known as 'spin' Chern numbers. This is the situation
in several well-known models \citep{kane_quantum_2005,bernevig_quantum_2006}
and for our case study of choice.}

\textcolor{black}{Just like before, we want to apply our NN to predict
band structures involving nontrivial unit-cell geometries. Starting
from any homogeneous (bulk) topological model, we can vary the underlying
parameters spatially in a periodic fashion, giving rise to a topological
metamaterial. To keep things simple and comparable to other examples,
we choose only two different values for the parameter in question
(similar to the potential landscape discussed so far).}

\textcolor{black}{Specifically, we consider a minimal model of a two-component
topological metamaterial -- a model with many intriguing features
that are interesting in their own right, not only as a benchmark for
our approach. In this metamaterial, we combine spatial regions of
trivial (spin Chern number ${\cal C}=0$) and topological (spin Chern
number ${\cal C}\neq0$) bulk material. Both regions are assumed to
share the same band gap of width $2m$ cf Fig. \ref{fig:Chern}(a).
The two materials are arranged according to a random unit-cell geometry
with four-fold rotational symmetry and lattice constant $a$, cf Fig.
\ref{fig:Chern}(b). The mismatch in Chern numbers at the closed domain
walls separating the two types of regions will give rise to topological
excitations inside the homogeneous bulk band gap, cf Fig. \ref{fig:Chern}(a).
The edge channels propagating along the domain walls themselves form
a lattice of closed loops. An isolated loop would have a discrete
eigenspectrum. These excitations will propagate chirally about any
given closed domain wall but also tunnel to adjacent domain walls,
cf Fig. \ref{fig:Chern}(c). This will give rise to the band structure
that we set out to investigate.}

\textcolor{black}{We model each of the two component materials using
the Bernevig-Hughes-Zhang model \citep{bernevig_quantum_2006}. This
model is known to capture well the physics of HgTe/CdTe semiconductor
quantum wells \citep{konig_quantum_2007}. For these materials, the
valence and the conduction bands have minimal splitting at the $\Gamma$
point, cf Fig. \ref{fig:Chern}(a). At this high-symmetry point, the
Bloch-waves are eigenstates of the quasi-angular momentum and are,
thus, orbital-polarized, cf the labels in Fig. \ref{fig:Chern}(a).
This allows to define the mass or band gap parameter $M$ as half
of the energy difference between the two Bloch-waves, $M=\pm m$.
The sign of $M$ is determined by the ordering of the bands and a
change of ordering is accompanied by a change of Chern number (by
one unit). In our conventions (see Appendix \ref{sec:Details-of-the-Spin_Hall}),
the Chern number is ${\cal C}=0,1$ for positive and negative mass,
respectively.}

\textcolor{black}{We model the composite metamaterial using the Bernevig-Hughes-Zhang
model but allowing a lattice-site-dependent mass $M(\mathbf{x})$
. Each random configuration $M(\mathbf{x})$ belongs to the $p_{4}$
wallpaper group (Lattice: Square. Point group: ${\cal C}_{4}$). Here,
the position $\mathbf{x}$ is defined on a 'microscopic' square lattice
of lattice constant $a/N$. The metamaterial unit cell contains $N\times N$
unit cells of the microscopic lattice and has lattice constant $a$.
This leads to a folding of the BZ giving rise to $N^{2}$ bands for
each band of the Bernevig-Hughes-Zhang model. Subsequent bands are
separated by local band gaps and we can assign to each band an integer
Chern number.}

\textcolor{black}{As we discussed above, we are primarily interested
in the band structure formed by the topological excitations in the
bandwidth of the single-domain band gap, cf Fig \ref{fig:Chern}(c-d).
The distribution of the Chern numbers for the first four bands above
the Fermi energy (here located in the middle of the whole band structure)
is shown in the stacked bar chart Fig \ref{fig:Chern}(f). We consider
the mesoscopic regime where the typical localization length $\xi$,
i.e. the transverse extent of the edge channels at the domain walls,
is larger than the microscopic lattice constant, but smaller than
the macroscopic one: $a/N\ll\xi\ll a$. Since $a$ sets the scale
for the typical distance between adjacent domain walls, we expect
the inter-domain hopping to be exponentially suppresed and, thus,
the bands to be well separated. In this regime, the band structure
is well approximated within a large-wave-length description encapsulated
in the Dirac equation
\begin{equation}
H_{D}=M(\mathbf{\hat{x}})\hat{\sigma}_{z}+v(\hat{k}_{x}\hat{\sigma}_{x}+\hat{k}_{y}\hat{\sigma}_{y}).\label{eq:Dirac_Ham}
\end{equation}
Here, the energy is counted off from the Fermi energy and $v$ is
the speed of the excitations. Moreover, $\hat{\sigma}_{i=x,y,z}$
is a set of Pauli matrices whose basis states are an $s$ and a $p_{-}$
orbital for $\sigma_{z}=1$ and $\sigma_{z}=-1$, respectively. Importantly,
the anti-unitary transformation $\Xi={\cal K}\sigma_{x}$ (where ${\cal K}$
is the complex conjugation) is a particle-hole symmetry, $\Xi^{-1}H_{D}\Xi=-H_{D}$.
Since $\Xi^{2}=1\!\!1$, our family of metamaterials is in the symmetry
class D. }

\textcolor{black}{We note that the assumption that the Pauli matrices
are defined on a specific basis of atomic orbitals affects only the
symmetry labels but not the band structure or the Chern numbers and,
thus, does not imply any loss of generality. In this sense Eq. (\ref{eq:Dirac_Ham})
and, thus, all results presented here go beyond our specific microscopic
model. This includes (but it is not limited to) scenarios in which
the spin is not conserved (but the mirror out-of-plane transformation
$M_{z}$ is a symmetry) and/or the particle-hole-symmetry is an emergent
symmetry not present in the microscopic model, see Appendix \ref{sec:Details-of-the-Spin_Hall}. }

\textcolor{black}{We have trained 13 NNs to predict the eight bands
around the Fermi energy and the corresponding Chern numbers for $\xi=v/M=a/10$
(after a trivial rescaling of the energy this is the only free parameter
in the large-wavelength description). We use the same method described
in Section \ref{sec:Training}. However, since we aim to predict the
band structure in the middle of the spectrum (instead of starting
from the minimal energy), we now face an additional challenge: given
that the symmetry-enhanced TB model and the original lattice model
have a different number of bands it is not clear which band should
correspond to which (see Appendix \ref{sec:Details-of-the-Spin_Hall}).}

\textcolor{black}{All of the NNs trained with this approach perform
well for most of the geometries, even without postselection, both
in the prediction of the symmetry labels and of the Chern numbers.
This includes even cases where the training grid (a $9\times9$ k-space
grid covering 1/4th of the BZ) would be too coarse to calculate the
Chern numbers as the sum of the Berry fluxes across the BZ \citep{fukui_chern_2005}.
For the purpose of validating the Chern numbers after the training,
we use a fine $62\times62$ k-space grid, to obtain reliable results.
We further note in passing that for the $p_{4}$ wallpaper group the
symmetry labels of a band determine whether the Chern number is odd
or even \citep{fang_bulk_2012}. In this setting, the parity of the
(spin) Chern number can be identified with the $\mathbb{Z}_{2}$ topological
invariant \citep{hasan_colloquium:_2010} which is, thus, symmetry-indicated.
However, this still leaves ample space for error in predicting the
Chern numbers. Consider, for instance, that the odd Chern numbers
$1$ and $-1$ are both likely to occur, cf distribution in Fig \ref{fig:Chern}(f).
Therefore, the good performance of the NNs on this task is non-trivial
and not merely enforced by symmetry. Due to the natural correlations
between band structure and Berry curvature, even the Berry curvature
predicted by the NN is most of the times qualitatively correct, cf
Fig \ref{fig:Chern}(f). This is remarkable because the NNs have not
received any information regarding the Bloch waves away from the high-symmetry
points. }

\textcolor{black}{We compare the predictions by different NNs by plotting
in a scatter plot the rms band structure deviation (radius of the
circles) and the fraction of correctly predicted Chern numbers ($y$-axis)
and symmetry fingerprints ($x$-axis), cf Fig. \ref{fig:Chern}(g).
Below, we refer to the latter two quantities as Chern and fingerprints
accuracy, respectively. Remarkably, all NNs perform at a similar level
for the rms band structure deviation and the fingerprint accuracy,
while they can be roughly divided into two groups when also the Chern
accuracy is taken into account. For the first group, the Chern accuracy
is significantly lower than the fingerprints accuracy. This reflects
that, as it should be expected for a non-symmetry-indicated topological
invariant, for a significant number of cases the Chern number is wrong
even though the symmetry fingerprints are correct. On the other hand,
for the other much larger group (11 out of 13 NNs) the two figures
almost perfectly coincide and assume a value larger than $95$\% {[}cf
red circles in Fig. \ref{fig:Chern}(g){]}. Thus, by postselecting
any of the NNs of the second group, we obtain a high Chern accuracy. }

\textcolor{black}{As anticipated above, the remaining errors are mostly
due to band gaps that are too small to be reliably resolved by the
NN. This is confirmed by recalculating the Chern- and fingerprint
accuracy, now taking into account only the bands separated from neighboring
bands by a minimal splitting larger than a small threshold ($0.01m$,
or roughly three standard deviations of the band structure rms deviation).
In this case, both accuracies are very close to $100$\% (the residual
error is at the level of around $1\permil$), cf the blue circle in
Fig. \ref{fig:Chern}(g). This proves that the accuracy of our Chern
number predictions after NN postselection are only limited by the
band structure precision (a similar conclusion will then hold also
for the symmetry labels.)}

\begin{figure}
\includegraphics[width=1\columnwidth]{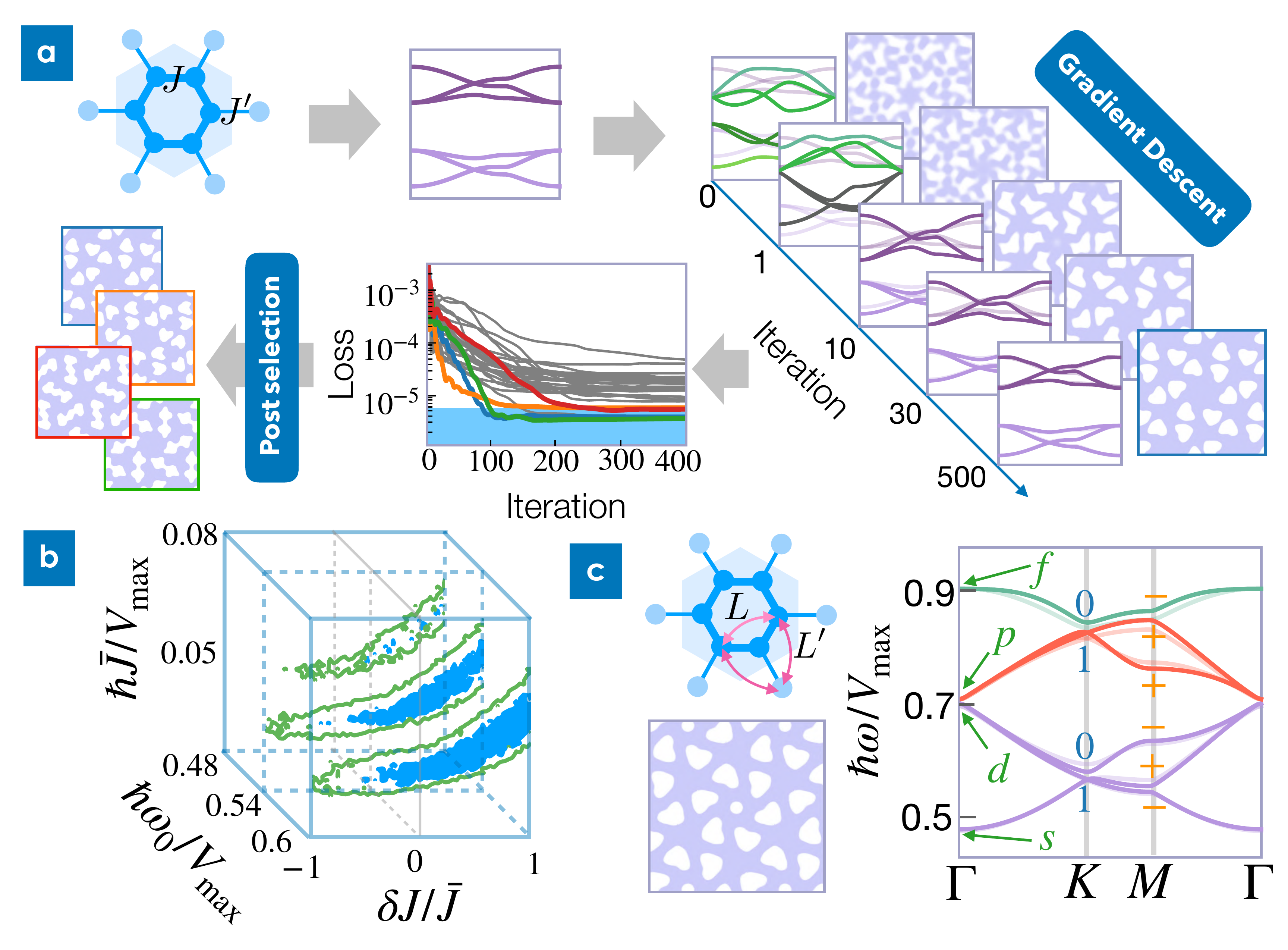}

\caption{\label{fig:Optimization} Using the NN for gradient-based optimization.
(a) Pipeline for finding a physical implementation of a given TB model
(unit cell shown). The band structure of the model is provided as
a target. A randomly initialized geometry is evolved until the NN-predicted
band structure (dark lines) approaches the target (bright lines).
We select those random trials that end up at a loss comparable to
the NN accuracy itself (blue shaded region). (b) Physically accessible
regions of the TB model parameter space: contours show the minimal
loss, achieved after $50$ trials, as a function parameters (onsite
energy $\omega_{0}$, average hopping $\bar{J}=(J+J')/2$, hopping
difference $\delta J=J'-J$). In the blue region, the minimal loss
is lower than the network accuracy. (c) Solutions for a more general
model, including also next-nearest neighbor hopping, supporting fragile
topological phases, see Appendix \ref{sec:Fragile-Tight-binding-model}
for more details. The parameters are $\hbar\omega_{0}=0.6V_{{\rm max}}$,
$\hbar J'=0.038V_{{\rm max}},$ $\hbar J=0.035V_{{\rm max}},$ $\hbar L'=0.002V_{{\rm max}},$
$\hbar L=-0.002V_{{\rm max}}$.}
\end{figure}

\section{Optimization }

Gradient-based optimization search for a geometry that maximizes some
reward is a powerful but numerically intensive design tool for photonic
devices \citep{cox_band_2000,men_robust_2014,nanthakumar_inverse_2019,christiansen_topological_2019}.
The numerical effort involved in calculating a large number of FEM
simulations represents a substantial bottleneck for explorative designs.
NNs offer a natural way out of this as it has been demonstrated in
a handful of pioneering works \citep{pilozzi_machine_2018,asano_optimization_2018,peurifoy_nanophotonic_2018,asano_iterative_2019}.
In contrast to these works, our approach allows to search for an arbitrary
input geometry. As explained above, this geometry is parametrized
via the Fourier coefficients of a smooth field that is then discretized
via a sigmoid function (see Appendix \ref{sec:OptimizationApp}).

An important goal consists in solving the inverse problem, where we
try to reach a given target band structure. This might be used, for
example, to find a physical implementation of some TB model of interest
(sharing the goal of \citep{matlack_designing_2018}), under the given
experimental constraints.

In Fig.~\ref{fig:Optimization}, we illustrate the procedure for
a TB model \citep{wu_topological_2016} that underlies fruitful applications
in topological photonics \citep{barik_topological_2018,parappurath_direct_2018}
and phononics \citep{brendel_snowflake_2018,cha_experimental_2018}.
The presence of local minima in the optimization landscape is easily
addressed by running multiple trials and post-selecting outcomes,
thanks to the 1000-fold acceleration produced by the NN.

We observe that the optimal geometry is not defined uniquely (Fig.~\ref{fig:Optimization}a),
since we only demand a match in the first few bands. This could be
exploited to select for structures that are easy to fabricate. Conversely,
however, it is not generally possible to reach arbitrary band structures,
due to physical constraints like the allowed values of the potential
(the refractive index contrast in the photonic case) and the unit
cell size. To delineate the accessible regions of the TB model parameters,
a scan with repeated optimization runs is required. Doing this for
a 3D parameter space (Fig.~\ref{fig:Optimization}b) even on a coarse
grid, the number of evaluations runs in the millions (Appendix \ref{sec:OptimizationApp}),
which does not present a problem for the NN but would be very impractical
otherwise. The resulting map can be used as a starting point for realizing
extended TB models, e.g. implementing fragile topological phases with
next-nearest neighbor hopping (Fig.~\ref{fig:Optimization}c).

Other reward functions can be used to optimize only for specific feature
combinations (like band gaps, group velocities, selected band representations,
etc.). More generally, one might even optimize potential landscapes
-- where smooth geometry deformations in real-space lead to some
band structure evolution that (e.g.) produces edge states with desired
properties. One important point in optimization is that the network
should give reliable robust predictions even away from training examples.
Empirically, this seems to be the case here, in our observations.
Nevertheless, this could be the domain of further study, possibly
exploiting the concept of adversarial approaches (where one tries
to slightly change the input in a deliberately disadvantageous way,
to maximize the deviation from the correct output; see e.g. \citep{jiang_vulnerability_2019}).

\section{Outlook }

The tight-binding network approach introduced here can be directly
applied to many other situations. These include, without any alterations
in the NN, finite-element calculations for electromagnetic and elastic
waves (where the execution speed advantage of the NN is enhanced by
further orders of magnitude). Moreover, direct extensions allow to
address band structures for metamaterials with inhomogeneous dissipation
and amplification (with complex eigenfrequencies and exceptional point
physics in reciprocal space), and driven nonlinear photonic crystals
or optomechanical arrays (where excitation pair creation leads to
a symplectic Hamiltonian structure and novel topological features).
Interactions on the mean-field level can be addressed as well, e.g.
using solutions of the Gross-Pitaevskii equation for matter waves
in optical lattices, or using density-functional theory results for
real materials (where the input could be atomic positions instead
of geometries, using the ideas of SchNet \citep{schutt_schnet_2018}).
We expect approaches like the one exemplified here to become a standard
part of the toolbox for metamaterial design.

\section*{Acknowledgments }

This work was supported by the European Union\textquoteright s Horizon
2020 Research and Innovation program under Grant No. 732894, Future
and Emerging Technologies (FET) - Proactive Hybrid Optomechanical
Technologies (HOT). We thank Leopoldo Sarra for fruitful feedback.

\appendix

\section{Network layout\label{sec:Network-layout}}

\begin{figure}
\includegraphics[width=1\columnwidth]{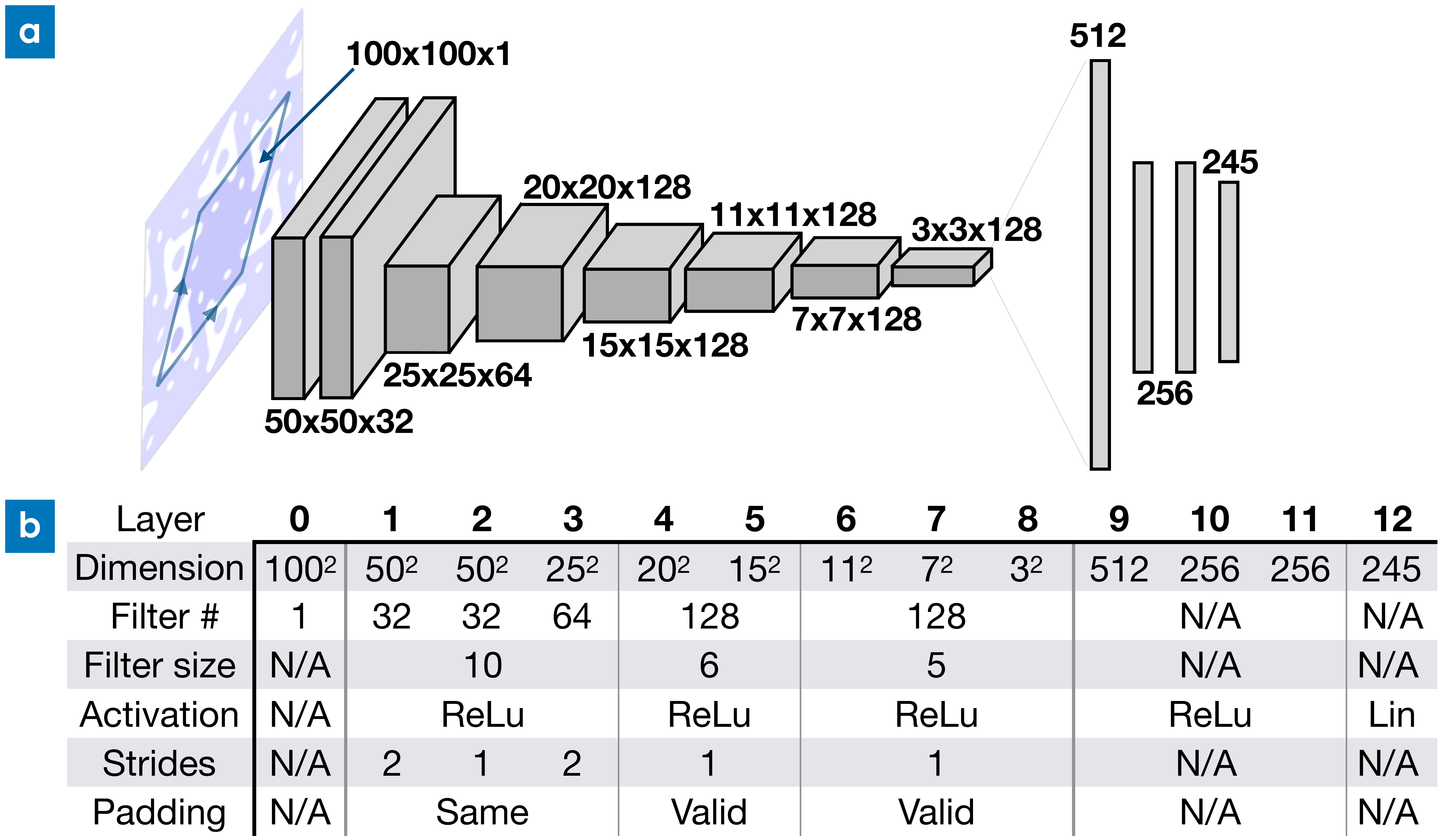}\caption{\label{fig:NNlayout}Details of the NN layout. (a) Sketch of the layout
of the neural network. The input image is the unit cell of the potential
(the region within the blue contour). The numbers indicate the dimensions
of the layers, where for the convolutional the last value is the number
of filters. The image is flattened before it is processed by the dense
layers. The final dense layer has 245 output neurons encoding the
independent coefficients of the TB model. (b) Detailed list of the
parameters for the network. 'Layer 0' labels the input image.}

\end{figure}

Here, we describe the layout of our NN. The network maps a two-dimensional
$L\times L$ image $V(\cdot)$, representing the potential inside
a (parallelogram-shaped) unit cell, onto a finite set of coefficients
of the symmetry enhanced TB model, see Appendix \ref{sec:Symmetry-enhanced-tight-binding}.
In our explorations, we have found that a good choice for the number
of convolutional layers as well as the kernel sizes is important for
robust and successful training, even though the fine details do not
matter. The proper structure depends mainly on the size of the input
potential. For the case of potentials with size $100\times100$, we
have found the layout described below to give good results. The layout
is sketched in Fig. \ref{fig:NNlayout}(a). A detailed list of parameters
is given in the table in Fig. \ref{fig:NNlayout}(b). The entire implementation
is using the TensorFlow framework.
\begin{itemize}
\item \textbf{Multi-layer convolutional network}: The first 8 layers of
the neural network are \lstinline!conv2D! layers with ReLU as activation
function. One goal of applying successive convolutional layers is
the reduction of the image size, which is usually done using Pooling
layers. In our case we instead reduce the image size mainly by using
\lstinline!stride=2! in some layers. The combination of the options
\lstinline!stride=2! and \lstinline!padding=same! leads to the reduction
of the pixels by a factor of four (a factor of two in each direction).
The combination \lstinline!stride=1! and \lstinline!padding=valid!
leads also to a reduction of the pixels by eliminating grid points
close to the boundaries (when the kernel is not completely within
the image). 
\item \textbf{Multi-layer fully connected network part}: after flattening
the result of the last \lstinline!conv2D! layer, 4 dense layers are
applied with \lstinline!dropout(0.15)! between each pair of dense
layers. The first 3 dense layers use also ReLU activations, while
the last dense layer uses linear activation to allow rescaling. The
numbers neurons are: 512/256/256/245. The number of neurons in the
last layer (245) corresponds to the number of independent coefficients
of our symmetry enhanced TB model, see Appendix \ref{sec:Symmetry-enhanced-tight-binding}.
\end{itemize}
The resulting output of the neural network is then interpreted as
the coefficients of the (symmetry-enhanced) tight-binding model (see
section above), which is diagonalized numerically. We run everything
except the eigenvalue calculations on the GPU. However, due to the
implementation of the diagonalisation in TensorFlow, it is important
to run \lstinline[language=Python]!tf.linalg.eigh! on the CPU instead
of the GPU. Otherwise the diagonalisation will take about 2 orders
of magnitude longer due to parallelization overhead.

Future improvements might include implementing \lstinline!conv2D!
layers with periodic boundary conditions, and implementing the convolution
operations on the actual mesh inside the parallelogram-shaped unit
cell (for a triangular lattice this would be a triangular mesh, instead
of the square mesh assumed in the tensorflow implementation). However,
as far as we could observe, these details did not prevent the network
from reaching a very good performance.

\section{Gradient descent for the combination of neural network and tight-binding
model\label{sec:Gradient-descent-for}}

One of the unconventional parts of our ansatz is the use of known-operator
learning, i.e. having numerical diagonalisation be part of the overall
pipeline leading from geometry to band structure. To be able to perform
gradient descent on this combination, we implemented a suitable modification
of the cost function introduced above. The main idea is to exploit
perturbation theory to obtain the derivative of the eigenvalues of
a matrix with respect to its coefficients, and to feed this analytical
expression into the tensorflow backpropagation pipeline.

The expression for the modified cost function can be derived by starting
with the derivative of the original cost function:

\begin{align*}
\frac{\partial C(\theta)}{\partial\theta}= & \Biggl<\sum_{l}2\left(\omega_{n,\mathbf{F_{\theta}}(V(\cdot))}^{{\rm NN}}(\mathbf{k})-\omega_{n}^{{\rm true}}(\mathbf{k})\right)\\
 & \times\frac{\partial\omega_{n,\mathbf{F_{\theta}}(V(\cdot))}^{NN}(\mathbf{k})}{\partial F_{\theta}^{(l)}(V(\cdot))}\,\frac{\partial F_{\theta}^{(l)}(V(\cdot))}{\partial\theta}\Biggr>_{V(\cdot),\mathbf{k},n}\,,
\end{align*}
where $F_{\theta}^{(l)}(V(\cdot))$ is determined by a neuron of the
output layer. Here we have been very careful in spelling out all the
dependencies; in particular, the band structure depends on the neural
network parameters $\theta$, via the tight-binding coefficients.

We can use the Rayleigh-Schrödinger perturbation-theory relation

\[
\frac{\partial\omega_{n,\mathbf{F_{\theta}}(V(\cdot))}(\mathbf{k})}{\partial F_{\theta}^{(l)}(V(\cdot))}=\frac{1}{\hbar}\left\langle \phi_{n}(\mathbf{k})\left|\frac{\partial\hat{H}_{\mathbf{k}}(\mathbf{F_{\theta}}(V(\cdot))}{\partial F_{\theta}^{(l)}}\right|\phi_{n}(\mathbf{k})\right\rangle 
\]
to calculate the derivative of the eigenvalues. Since $\hat{H}_{\mathbf{k}}(\mathbf{F_{\theta}}(V(\cdot)))$
is linear in every coefficient $F_{\theta}^{(l)}(V(\cdot))$, $\frac{\partial\hat{H}_{\mathbf{k}}(\mathbf{F_{\theta}}(V(\cdot)))}{\partial F_{\theta}^{(l)}}$
is a numerical constant. Therefore, we must calculate this expression
only once before training and can then use it for all training steps.
$F_{\theta}^{(l)}(V(\cdot))$ is also independent of the sum over
the $\mathbf{k}$ and $n$, which is why we can rewrite the derivative
of the cost function as

\[
\frac{\partial C(\theta)}{\partial\theta}=\left\langle \sum_{l}v_{l}\,\frac{\partial F_{\theta}^{(l)}(V(\cdot))}{\partial\theta}\right\rangle _{V(\cdot),\mathbf{k},n}\,,
\]
with

\begin{align*}
 & v_{l}(V(\cdot),\mathbf{k},n)=\frac{2}{\hbar}\left(\omega_{n,\mathbf{F_{\theta}}(V(\cdot))}^{{\rm NN}}(\mathbf{k})-\omega_{n}^{{\rm true}}(\mathbf{k})\right)\\
 & \times\left\langle \phi_{n,\mathbf{F_{\theta}}(V(\cdot))}(\mathbf{k})\left|\frac{\partial\hat{H}_{\mathbf{k}}(\mathbf{F}_{\theta}(V(\cdot)))}{\partial F_{\theta}^{(l)}}\right|\phi_{n,\mathbf{F_{\theta}}(V(\cdot))}(\mathbf{k})\right\rangle \:.
\end{align*}
Hence, we arrive at the conclusion that we can use
\begin{align}
C^{{\rm NN}}(\theta) & =\left\langle \sum_{l}v_{l}(V(\cdot),\mathbf{k},n)\,F_{\theta}^{(l)}(V(\cdot))\right\rangle _{V(\cdot),\mathbf{k},n}\nonumber \\
 & =\left\langle \mathbf{v(}V(\cdot),\mathbf{k},n\mathbf{)\,\cdot F_{\theta}(}V(\cdot)\mathbf{)}\right\rangle _{V(\cdot),\mathbf{k},n}\label{eq:CostSM}
\end{align}

as cost function for the neural network.

Indeed, the $\theta-$gradient of this cost function is the same as
for the original one, as long as we postulate that \textbf{$\mathbf{v}$}
is to be treated as independent of $\theta$. Since $\mathbf{v}$
is a vector with the number of coefficients as number of entries,
this cost function is realised in TensorFlow as a simple scalar product
between the output layer of the NN and the vector $\mathbf{v}$. Each
training step of the neural network consists of the calculation of
$\mathbf{v}$ (for a batch of training samples) and the usual gradient
descent applied to the cost function $C^{NN}$.

\section{Symmetry-enhanced tight-binding Hamiltonian\label{sec:Symmetry-enhanced-tight-binding}}

Here, we give more details on the 'symmetry enhanced' tight-binding
models whose parameters are predicted by our NN and subsequently used
to calculate the band structures. {[}There is one such TB model for
each symmetry group considered (space group plus time-reversal symmetry
when applicable.)

The challenge in defining such a TB models is that they should be
able to reproduce the low energy bands of a broad distribution of
potentials. Moreover, the number of underlying orbitals and parameters
should remain as small as possible to keep the diagonalization of
the TB Hamiltonian numerically inexpensive. 

\subsection*{Wallpaper group $\boldsymbol{p_{6}}$ with time reversal symmetry}

In order to estimate how large the Hilbert space of our TB model should
be, we define the 'occupation' $n_{\xi}^{(V(\cdot))}$ in the lowest
seven bands for the potential $V(\cdot)$ and the irreps $\xi$, $\xi=s,p,d,f,0,1,+,-$.
The number of orbital required for our TB model will then depend on
the maximal occupations $n_{\xi}^{({\rm max})}$ over all training
samples, $n_{\xi}^{({\rm max})}={\rm {\rm Max}_{V(\cdot)}n_{\xi}^{(V(\cdot))}}$. 

We have (somewhat arbitrarily) decided to build our TB model using
only orbitals localized about the ${\cal C}_{6}$ rotocenters. We
denote by $\tilde{n}_{{\rm l}}^{({\rm TB})}$ the number of $l$-orbitals,
$l=s,p,d,f$. The number of different orbitals in real space then
determines the number $n_{\xi}^{(TB)}$ of Bloch waves that are available
for each irrep. at the high symmetry points, e.g. $n_{0}^{({\rm TB})}=\tilde{n}_{s}^{({\rm TB})}+\tilde{n}_{f}^{({\rm TB})}$
and so on. Requiring $n_{{\rm \xi}}^{({\rm TB})}\ge n_{\xi}^{({\rm max})}$
for all irreps (such that for all samples enough Bloch waves with
the right symmetry are available) results in a lower bound on $\tilde{n}_{l}^{({\rm TB})}$.
For the $50000$ training samples used to train our NN we have found
$n_{\xi}^{({\rm max})}=3$, for $\xi=s$ and $n_{\xi}^{({\rm max})}=2$
otherwise. Accordingly, we have chosen to have $\tilde{n}_{s}^{({\rm TB})}=4$
$s$-orbitals, $\tilde{n}_{p}^{({\rm TB})}=4$ $p$-orbitals, $\tilde{n}_{d}^{({\rm TB})}=3$
$d$-orbitals, $\tilde{n}_{f}^{({\rm TB})}=3$ $f$-orbitals (well
above the lower bound set by $n_{{\rm max}}^{(\xi)}$.) 

We note that while all unperturbed orbitals for our TB model are localized
about the same Wyckoff position, the Wannier states for an isolated
set of bands can still be hybridized orbitals localized about different
Wyckoff positions. This may happen because the hoppings between different
TB orbitals can be larger compared to the typical onsite energy differences.
Thus, our choice of the Wyckoff position for the unperturbed orbitals
is akin to a choice of basis. We also restrict the hopping to nearest-neighbor
orbitals. This choice reduces the number of output neurons and, thus,
the overall complexity of the NN while still turning out to be adequate
to obtain a well trained NN in the examples considered in this work. 

Next we derive the explicit form of the TB model described above in
terms of the appropriate set of independent onsite energies and hoppings
amplitudes. The constraints imposed by the ${\cal C}_{6}$ symmetry
that connect hopping rates in different directions are most easily
taken into account using a basis of ${\cal C}_{6}$ symmetric Wannier
orbitals $\{|\tilde{W}_{n,m}\rangle\}$ where $n$ is the principal
quantum number and $m$ is the quasi-angularmomentum, $\hat{R}_{\pi/3}|\tilde{W}_{n,m}\rangle=e^{-im\pi/3}|\tilde{W}_{n,m}\rangle$
with $m=0,\pm1,\pm2,3$. In the corresponding basis of Bloch waves,
one can then easily add the contributions from all hopping directions
to find

\[
\hat{H}_{\mathbf{k};n,m;n'm'}(\mathbf{k)}/\hbar=\delta_{m,m'}\delta_{n,n'}\omega_{n,|m|}+\tilde{J}_{n,m;n',m'}f_{m-m'}(\mathbf{k})
\]
where $\omega_{n,|m|}$ are the onsite energies, $\tilde{J}_{n,m;n',m'}$
are the hopping amplitudes in the direction of the lattice vector
$\boldsymbol{a}_{1}=a(1,0)$, and the functions $f_{\Delta m}(\mathbf{k})$
are independent of the potential,
\begin{align*}
 & f_{0}(\mathbf{k})=\cos(\mathbf{k}\cdot\boldsymbol{a}_{\mathbf{1}})+\cos(\mathbf{k}\cdot\boldsymbol{a}_{\mathbf{2}})+\cos(\mathbf{k}\cdot\boldsymbol{a}_{\mathbf{3}}),\\
 & f_{1}(\mathbf{k})=-f_{-1}^{*}(\mathbf{k})=\\
 & -i[\sin(\mathbf{k}\cdot\boldsymbol{a}_{\mathbf{1}})+e^{-i\pi/3}\sin(\mathbf{k}\cdot\boldsymbol{a}_{\mathbf{2}})+e^{-i2\pi/3}\sin(\mathbf{k}\cdot\boldsymbol{a}_{\mathbf{3}})],\\
 & f_{2}(\mathbf{k})=f_{-2}^{*}(\mathbf{k})=\\
 & \cos(\mathbf{k}\cdot\boldsymbol{a}_{\mathbf{1}})+e^{i4\pi/3}\cos(\mathbf{k}\cdot\boldsymbol{a}_{\mathbf{2}})+e^{i2\pi/3}\cos(\mathbf{k}\cdot\boldsymbol{a}_{\mathbf{3}}),\\
 & f_{3}(\mathbf{k})=-i[\sin(\mathbf{k}\cdot\boldsymbol{a}_{\mathbf{1}})-\sin(\mathbf{k}\cdot\boldsymbol{a}_{\mathbf{2}})+\sin(\mathbf{k}\cdot\boldsymbol{a}_{\mathbf{3}})],
\end{align*}
with $\boldsymbol{a}_{2}=a(1,\sqrt{3})/2,$ $\boldsymbol{a}_{3}=a(-1,\sqrt{3})/2$.
Due to the time-reversal symmetry the onsite energy is the same for
states with equal principal quantum number $n$ and opposite quasi-angular
momentum $m$. This, results in $\tilde{n}_{s}^{({\rm TB})}+\tilde{n}_{{\rm f}}^{({\rm TB})}+\tilde{n}_{{\rm p}}^{({\rm TB})}+\tilde{n}_{{\rm d}}^{({\rm TB})}$
real independent onsite energies which are represented by an equal
number of output neurons $F^{(l)}$, cf Eq. (\ref{eq:CostSM}). The
hopping amplitudes $\tilde{J}_{n,m;n',m'}$ are also constrained by
the symmetries of the problem. The relevant constraints are most easily
expressed by switching to a time-reversal invariant basis of Wannier
states $\{|W_{n,l}\rangle\}$, $l=s,p_{1},p_{2},d_{1},d_{2},f$, where
$|W_{n,s(f)}\rangle=|W_{n,0(3)}\rangle$, and

\begin{align}
|W_{n,p(d)_{1}}\rangle & =\frac{1}{\sqrt{2}}\left(|\tilde{W}_{n,1(2)}\rangle+|\tilde{W}_{n,-1(2)}\rangle\right)\nonumber \\
|W_{n,p(d)_{2}}\rangle & =\frac{-i}{\sqrt{2}}\left(|\tilde{W}_{n,1(2)}\rangle-|\tilde{W}_{n,-1(2)}\rangle\right).\label{eq:defpxpy}
\end{align}
{[}Here, we have also implicitly fixed the sum of the phases of states
with equal $n$ and opposite quasi-angular momentum $m$ by assuming
${\cal T}|\tilde{W}_{n,m}\rangle=|\tilde{W}_{n,-m}\rangle$.{]} Because
of the time-reversal symmetry, the hopping amplitudes $J_{n,l;n',l'}$
in the time symmetric basis are real. Moreover, using the ${\cal C}_{2}$
symmetry and that the Hamiltonian should be hermitian one finds the
additional constraint
\[
J_{n,l;n',l'}=\pm J_{n',l';n,l}
\]
 where the positive sign applies when both orbitals have the same
behavior (odd or even) under the ${\cal C}_{2}$ symmetry and the
negative sign applies otherwise, e.g. $+$ when $l=s$ and $l'=d_{1(2)}$
(both orbitals are even) and $-$ for $l=s$ and $l'=f$ ($s$ is
even while $f$ is odd). Taking into account these additional constraints,
there are $(N_{{\cal H}}^{2}+N_{{\cal H}})/2$ real independent hopping
amplitudes, $N_{{\cal H}}=\tilde{n}_{s}^{({\rm TB})}+\tilde{n}_{{\rm f}}^{({\rm TB})}+2(\tilde{n}_{{\rm p}}^{({\rm TB})}+\tilde{n}_{{\rm d}}^{({\rm TB})})$.
These are represented by the same number of output neurons $F^{(l)}$,
cf Eq. (\ref{eq:CostSM}). 

\section{Generating Training Samples\label{sec:Generating-Training-Samples}}

When training a network on simulation results, an arbitrary random
distribution of training samples can in principle be chosen. However,
for best accuracy it is beneficial to have these samples be as close
as possible (statistically) to typical use cases encountered in later
applications.

We start by generating a periodic smooth 2D random Gaussian field,

\begin{equation}
\phi(\mathbf{x})=\sum_{\mathbf{\mathbf{k}}}A(\mathbf{k})e^{i\mathbf{\mathbf{k}}\cdot\mathbf{x}},\label{eq:periodicfunction}
\end{equation}
where the wavevectors $\mathbf{k}$ lie on the reciprocal lattice,
\begin{equation}
\mathbf{k}=n_{1}\boldsymbol{b}_{1}+n_{2}\boldsymbol{b}_{2}.\label{eq:reciplatt}
\end{equation}
Here, $n_{1}$ and $n_{2}$ are integers and $\boldsymbol{b}_{1}$
and $\boldsymbol{b}_{2}$ are reciprocal lattice vectors.

The Fourier coefficients $A(\mathbf{k})$ respect the underlying symmetry
(again, in our chosen example, they are symmetric under 60-degree
rotations). Otherwise, they are complex Gaussian-distributed random
numbers (of zero mean), with variance

\begin{equation}
\left\langle \left|A(\mathbf{k})\right|^{2}\right\rangle =\frac{C}{\left|\mathbf{k}a\right|^{\alpha}}f(\mathbf{k})\,.\label{eq:FourierCompDist}
\end{equation}
The function $f(\mathbf{k})$ is $1$ for small $\left|k\right|$
and implements a cutoff for larger $k$. In our case we choose $C=2$
and set $f(\mathbf{k})=0$ for $n_{1(2)}>6$ , cf Eq. (\ref{eq:reciplatt}).
The exponent $\alpha$ determines how smooth the field appears (in
our case, $\alpha=1$). \textcolor{black}{After training we have checked
that our NN still performs well for random validation potentials drawn
from a similar distribution but with a substantially larger cut-off.
This shows that the band structure for the low energy bands is insensitive
to fine details of the potential (on a length scale smaller than the
one set by our cut-off). This makes sense because the underlying Bloch
waves should remain smooth to reduce the kinetic energy.}

The examples treated in the main text are inspired by photonic or
phononic crystals, where two materials only are involved. This means
we want to provide a ``digitized'' potential, starting from the
smooth field $\phi$. That is achieved by the help of the rounded
step function, the sigmoid $\sigma(x)=1/(1+e^{-x})$:

\begin{equation}
V(\mathbf{x})=V_{{\rm max}}\sigma(\beta\phi(\mathbf{x}))\,,\label{eq:Beta_def}
\end{equation}
where smaller $\beta$ imply a more gradual step. For the training
we used sharp step functions corresponding to the limit $\beta\to\infty.$

\section{Training of the neural network\label{sec:Training-of-the}}

To train the neural network, we use 50,000 samples of random potentials,
with the correct band structure evaluated at 79 points. These points
are evenly distributed within one sixth of the Brillouin zone.

Out of these 50,000 samples, 1024 are reserved for calculating the
validation loss and the remaining 48,976 samples are used for training.
We use the widespread Adam optimizer, with parameters \lstinline!Adam(lr=0.0001, epsilon=10e-8)!.
The dropout rate between the dense layers is chosen to be 0.15. The
training spans many epochs ($\sim1000$). Before every training epoch
we reshuffle the training samples.

We recall that (as mentioned in the main text) in order to train the
NN also on the symmetry of the Bloch waves at the high-symmetry points,
we add to the \emph{global} (comprising an average over the BZ) cost
function Eq. (\ref{eq:cost}) other \emph{local} terms for each Block
of the Hamiltonian at each high symmetry point. Each block corresponds
to an irrep. of the proper group of the relevant high symmetry point
and the corresponding \emph{local} cost function term has (except
for the average over the BZ) the same form as Eq. (\ref{eq:cost}).
The overall cost function will then be a weighted sum of the \emph{global}
and the \emph{local} cost functions with the weight ratio $r$ between
the \emph{local} and the \emph{global} contributions playing an important
role during training. 

\textcolor{black}{Since the }\textcolor{black}{\emph{global}}\textcolor{black}{{}
cost function is oblivious of the symmetry labels, it tends to prevent
a change in the ordering of the bands. Thus, the }\textcolor{black}{\emph{local}}\textcolor{black}{{}
cost functions should dominate the global cost function at least until
the fraction of correct symmetry labels is high enough (indicating
that, for the overwhelming majority of the geometries, the bands are
correctly ordered). On the other hand, a too large weight ratio $r$
might imply an excessive emphasis on the high-symmetry points and,
thus, can be detrimental to the overall quality of the band structure
predictions. We note in passing that $r\sim1$ is to be regarded as
comparatively large because, assuming (to fix the ideas) that the
band structure deviations are of the same order across the BZ, it
implies that the contribution to the overall training gradient from
a single high-symmetry point (via the }\textcolor{black}{\emph{local}}\textcolor{black}{{}
cost functions) would be similar as the combined contributions from
all over the BZ (via the }\textcolor{black}{\emph{global}}\textcolor{black}{{}
cost function). Indeed, we have empirically observed that using a
constant $r$ of the order $r\sim1$ during the whole training run
produces accurate NNs and that one can improve even further the NNs
by decreasing $r$ toward the end of a training run. As it is often
the case for supervised learning, the training is limited by the onset
of overfitting. Good results are already obtained on a time scale
of $\sim100$ epochs but the onset of over-fitting occurs only on
a time scale of $\sim1000$ epochs. {[}Thus, it is worth to train
for $\sim1000$ epochs to obtain optimal accuracy.{]} The onset of
overfitting proves that our ansatz for the symmetry-enhanced TB model
(with all orbitals in the Wyckoff position $1a$ and nearest-neighbor
coupling, cf Appendix \ref{sec:Symmetry-enhanced-tight-binding})
does not represent a bottleneck for the achievable rms band structure
deviation.}

\textcolor{black}{We have observed that the training allows a large
degree of flexibility in the choice of the hyper-parameters, such
as optimizer, learning rate, batchsize, network layout, time-dependence
of the relative weight ratio $r$, etc. A good choice of hyper-parameters
consistently (for all random initial conditions of the network parameters)
leads to a low rms band structure deviation (of the order of $\sim0.001V_{{\rm Max}}$)
and a high fraction of correctly predicted symmetry labels. Nevertheless,
the predictions for different training runs might still be qualitatively
different, e.g. unlucky initial conditions may lead to spurious Dirac
cones or (for the two-component topological metamaterial case study,
cf Section \ref{sec:Strong-topological-phases}) to the wrong Chern
numbers. These wrong predictions are strongly correlated from sample
to sample. This means that they are either present for a significant
portion of validation samples (and are, thus, easy to detect) or are
not present at all. Thus, they can be easily eliminated by training
a few NNs and discarding the unreliable NNs, cf Section \ref{sec:Strong-topological-phases}.}

\textcolor{black}{Before the training of the NN, we pick a 'target'
number $N_{{\rm target}}$ of bands that we aim to predict. In this
work, we have shown results for different $N_{{\rm target}}$, $N_{{\rm target}}=4,6,8$.
In addition, we have performed numerical experiments with $N_{{\rm target}}=10$
for the $2$D Schrödinger equation, obtaining accurate predictions
($0.004V_{{\rm Max}}$ rms deviation and $99\%$ of correctly predicted
labels). An interesting question is how large can we increase $N_{{\rm target}}$.
We expect to be able to increase $N_{{\rm target}}$ somewhat above
$N_{{\rm target}}=10$ but that we will encounter a bottleneck in
the required number of trainable parameters for the NN. This scales
as the square of the number of independent parameters in our symmetry-enhanced
TB which itself scales as $N_{{\rm target}}^{2}$ (thus, overall we
have a quartic dependence). One could get around this problem by training
different NNs on different band numbers, e.g. one NN for the first
10 bands and a second for the next 10 bands. If this approach works,
the number of NN trainable parameters as a fuction of the overall
number of bands to be predicted will scale as $N_{{\rm target}}^{d+1/d}$
where $d$ is the dimension, e.g. $d=2$ in $2$D. This scaling is
governed by the number of potential grid points required to converge
to the continuum limit ($\propto$$N_{{\rm target}}$). This determines
both the number of neurons in each convolutional layer ($\propto$$N_{{\rm target}}$)
and the number of convolutional layers ($\propto$$N_{{\rm target}}^{1/d}$).}

Since the eigenvalue calculation is performed on the CPU, the duration
of 1 epoch depends strongly on the CPU. The workload on the CPU depends
on the number of ${\bf k}$-points in the global cost function. With
batchsize 16 and 79 reciprocal points, on a NVIDIA RTX 6000 and a
Xeon Gold 6130 with 16 cores, one epoch takes about 100 seconds.

\section{Accuracy for the network\label{sec:Accuracy-for-the}}

\textcolor{black}{The NN used to produce the results in Figures \ref{fig:Neural-network-based-prediction-},\ref{fig:WuandHu},\ref{fig:Topology},
and \ref{fig:Optimization}  has a rms band structure deviation of
$\approx0.0025V_{{\rm max}}$. This figure is calculated on a grid
with 821 grid points equally distributed within one sixth of the unit
cell (much finer than the training grid which had only 79 grid points).
As we noted above the rms deviation will be slightly different for
different training runs with equal training hyper-parameters (but
different initialization of the NN trainable parameters). Nevertheless,
it remains of a few $\permil$ of the overall scale of the band structure
(here $V_{{\rm Max}}$) for a wide range of training hyper-parameters
in all case studies investigated in this work. }

\begin{figure}
\includegraphics[width=0.7\columnwidth]{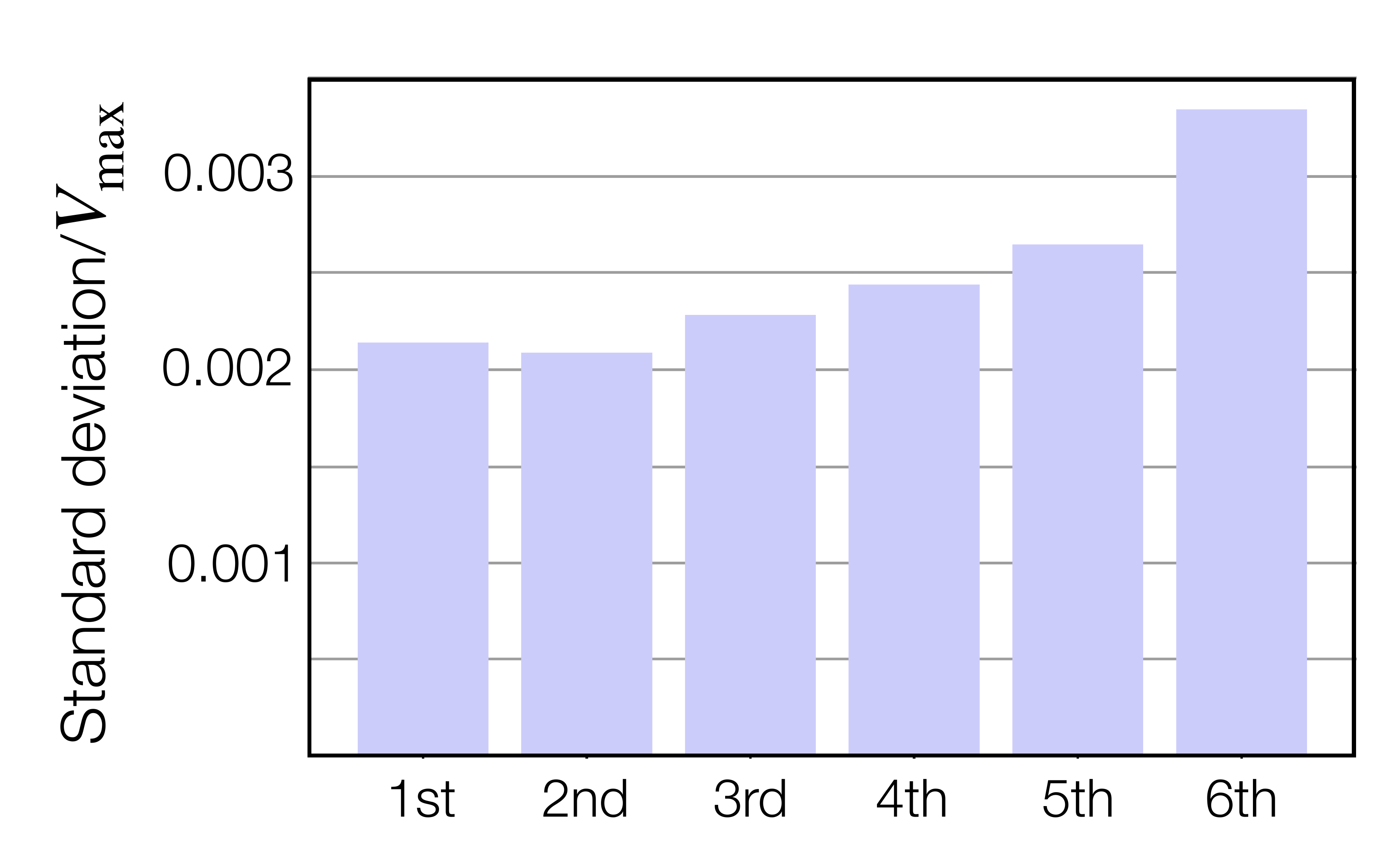}

\caption{\label{fig:accuracy}Rms deviation of the individual bands for the
NN used to produce Figures \ref{fig:Neural-network-based-prediction-},\ref{fig:WuandHu},\ref{fig:Topology},
and \ref{fig:Optimization} .}
\end{figure}

\textcolor{black}{We have empirically observed a slight trend for
increasing rms deviation for higher bands, cf Fig.~\ref{fig:accuracy}.
We attribute this to a higher sensitivity of higher energy bands to
the fine details of the potential. We have checked that this effect
can be compensated by increasing the number of training samples. }

If one wants to compare these deviations to the band gaps (as a natural
scale), we can, e.g. obtain the sample-average of the minimal band
gaps for disconnected bands, in which case the deviations represent
2.5\% of the bandgap value obtained in that way. On the other hand,
the sample-average of the \emph{$\mathbf{k}$-}averaged (not minimal)
band gaps for disconnected bands is slightly larger, yielding a relative
deviation of 2.1\%.

\section{Performance Gain\label{sec:Performance-Gain}}

\subsection{Speed advantage in predicting band structures }

One of several advantages of using a neural network for predicting
symmetries and band structures is the dramatically increased speed
of calculations vs. direct evaluations.

The performance gain offered by the neural network depends on the
algorithm which it replaces, as well as on the number of points in
reciprocal space, the structure of the neural network and the number
of bands which one is interested in.

For the results of this paper, the neural network should predict the
same band structure one would obtain by using the Schrödinger equation
on a periodic potential with 100$\times$100 grid points. In the absence
of a trained neural network, this would be accomplished for one ${\bf k}$-point
in the reciprocal space by calculating the eigenvalues of a sparse
10,000$\times$10,000 matrix, where the diagonal elements correspond
to values of the potential on the grid points.

For the numerical calculation we use for diagonalisation \lstinline!scipy.sparse.linalg.eighs!,
which uses the ``implicitly restarted Lanczos method'' to find the
eigenvalues and eigenvectors for the first 6 bands in our case. In
our comparison, we will count only the time needed to calculate the
eigenvalues for the case of the numerical method (which works in favor
of the numerical method). As reference hardware, both for the direct
numerical calculation and the neural network, a i5-6267U (2 cores,
4 threads, 2.9 Ghz) is used, which is as a typical mobile CPU. On
this hardware, the diagonalisation takes about 2-3 seconds for 3 points
and \textbf{80 seconds for the 79 points} which we also use for training.
On the other hand, the same task takes for the neural network \textbf{0.067
seconds for 79 points} (and 0.23 seconds for a much finer grid of
821 points).

This shows that the neural network performs much faster than the direct
Lanczos-based diagonalisation of the Schrödinger Equation does, even
for very few points. The advantage of the NN grows with the number
of points: note that the calculation time in the case of the neural
network can be split into the calculation of the coefficients for
the tight-binding model and the subsequent calculation of its band
structure (by diagonalization of a small matrix), where the former
is independent of the number of ${\bf k}$-points. Since the creation
of the tight binding model is written in python one could accomplish
further speed-up for the neural network.

\subsection{Overall performance gain }

As in all neural-network applications, there are two scenarios to
evaluate the cost-benefit and overall performance gain of this approach.

(i) The goal is to deploy the network for obtaining speedup on whatever
hardware is available (including, e.g., the type of cluster used for
training). In that case, the training effort needs to be accounted
for. Break-even will be reached when the network has been used to
accelerate band structure evaluations on a number of potentials that
is at least larger than the initial number of training samples. For
our approach, this is easily the case for the optimization of band
structures (as well as for large-scale statistical exploration and
random discovery).

(ii) The cost-benefit analysis turns out to be even more advantageous
when the explicit goal has been to deploy the network on modest computing
hardware (e.g. laptops operated by the end-users). In that case, the
cost of generating the training samples and performing the training
(on a cluster) need not be taken into account, since that hardware
by definition would not have been available to the end-user.

\section{Symmetry Fingerprints\label{sec:Symmetry-Fingerprints}}

Here, we give more details regarding the symmetry fingerprints used
to identify EBRs and topological bands in the main text. Equivalent
concepts are also presented in Refs. \citep{kruthoff_topological_2017}
and \citep{po_symmetry-based_2017}. The symmetry fingerprint of an
isolated set of bands\textbf{ }groups in a single $\mathbb{N}^{n}$
array all the information about the symmetry of the Bloch waves at
the maximal $\mathbf{k}$-points.

At each maximal $\mathbf{k}$-point, we define the 'occupation' $n_{\xi}$
as the number of (degenerate) orbitals belonging to each irrep $\xi$.
The occupation numbers $n_{\xi}$ are subject to linear constraints
known as compatibility relations \citep{po_symmetry-based_2017,bradlyn_band_2018}.
This reduces the number of independent 'occupations' to
\[
n={\rm number\,}{\rm of}\,{\rm irreps}-{\rm number\,}{\rm of\,}{\rm linear\,}{\rm constraints}
\]
The simplest compatibility relation is that the number of bands is
the same at all maximal $\mathbf{k}$-points. Another important example
of a compatibility relation is realized in crystals with mirror symmetry.
For each high symmetry line that is invariant under a mirror symmetry
of the crystal and connects two maximal $\mathbf{k}$-points, a compatibility
relation fixes\textbf{ }the numbers of states with a given parity
to be equal at the two maximal $\mathbf{k}$ points. Such compatibility
relations derived from mirror symmetry allow to predict connections
between bands that lie on\textbf{ }a high-symmetry line, based only
on the spectrum and irreps at the maximal $\mathbf{k}$ points \citep{po_symmetry-based_2017,bradlyn_topological_2017}.

For the $p_{6}$ group, the maximal ${\bf k}$-point are the high
symmetry points $\Gamma$, $K$, and $M$ and the respective proper
groups are the rotational groups ${\cal C}_{n}$, with $n=6$ for
$\Gamma$, $n=3$ for $K$, and $n=2$ for $M$. In this case, the
time-reversal-symmetric irreps are identified by the absolute value
of the quasi-angular momentum $|m|\leq n/2$. To avoid confusion,
here and in the main text we use $\xi=|m|=0,1$ to label the irreps
of ${\cal C}_{3}$, while using the atomic physics inspired labels
$\xi=s$, $p$, $d$ and $f$ in place of $|m|=0,1,2,3$ for the irreps
of ${\cal C}_{6}$, and the labels $\xi=+$ and $\xi=-$ in place
of $m=0$ and $m=1$ for the irreps of ${\cal C}_{2}$ (the normal
modes are either odd or even). We are, thus, left with $8$ occupation
numbers: $n^{(s)}$, $n_{p}$, $n_{d}$, and $n_{f}$ for the $\Gamma$
point; $n_{0}$, and $n_{1}$ for the $K$ point; $n_{+}$, $n_{-}$
for the $M$ point. However, taking into account that the overall
number of bands should be the same at all high-symmetry\textbf{ }points,
we find the compatibility relations, 
\begin{align}
n_{-} & =n_{s}+2(n_{p}+n_{d})+n_{f}-n_{+},\nonumber \\
n_{1} & =(n_{s}+2(n_{p}+n_{d})+n_{f}-n_{0})/2.\label{eq:CompNB}
\end{align}
Thus, all the information regarding the symmetry labels is grouped
in the six-dimensional array $(n_{s},n_{p},n_{d},n_{f},n_{0},n_{+}).$

\section{Details of the topological exploration\label{sec:AppTopexpl}}

\begin{figure}
\includegraphics[width=0.7\columnwidth]{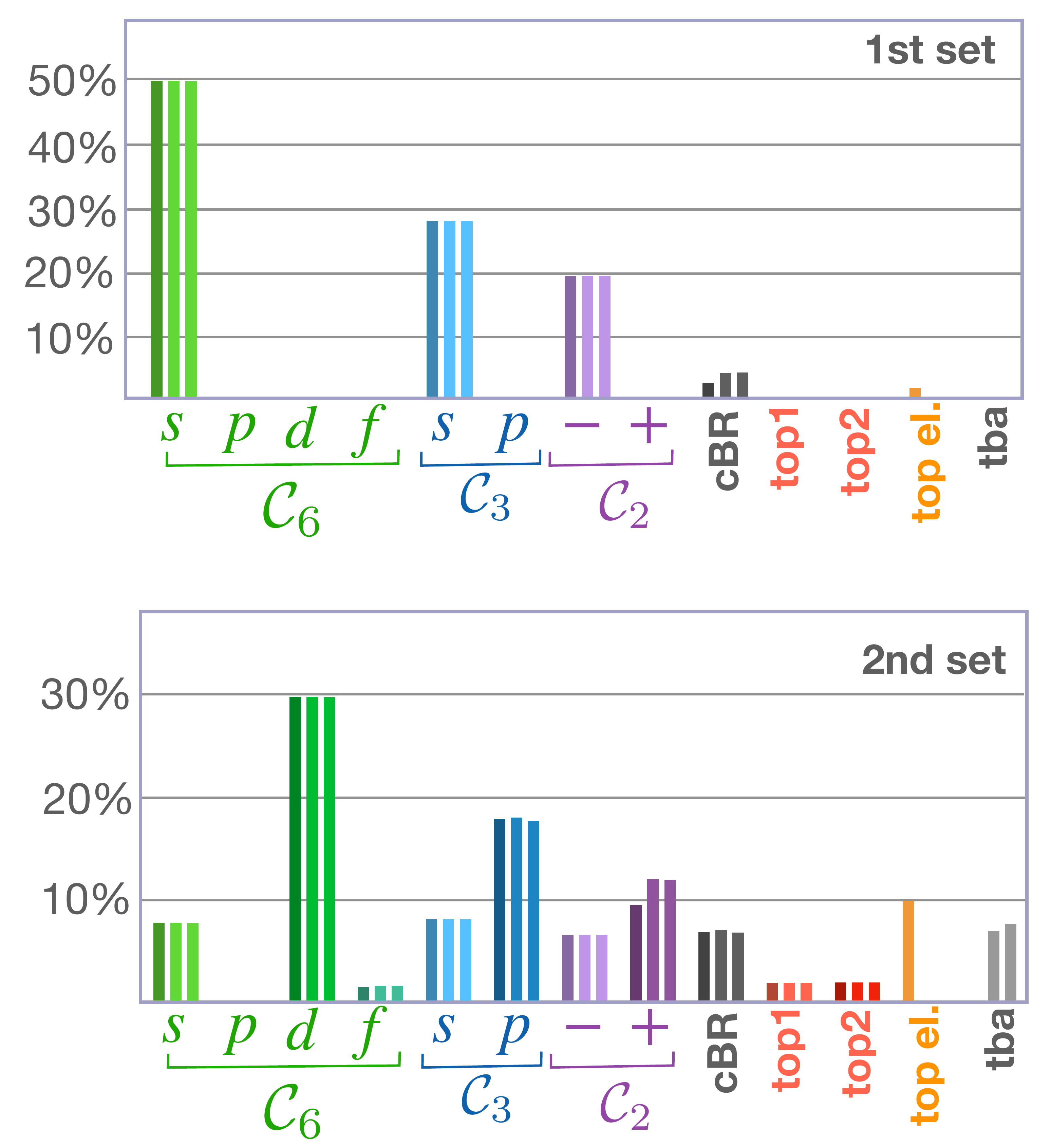}

\caption{\label{fig:add_results_statistics}(quasi-)BR distribution calculated
using three different methods to determine the bands connectivity.
The methods are (from left to right for each entry): (i) Standard
method of topological quantum chemistry which neglects connections
away from high symmetry points. (ii) Taking into account the connections
away from high-symmetry points which are predicted by the parity compatibility
relation. (iii) Looking for $\pi$-defects in the Berry flux. This
method is able to detect all connections (if a fine enough grid is
used). The column top1 and top2 refer to the two fragile topological
phases discovered by our NN, cf. Fig. 3 of the main text. The standard
method would predict other types of topological bands (top. el.).
These sets of bands turn out not to be isolated once the connections
away from the high-symmetry points are taken into account with the
other two methods. All results have been calculated using the first
six bands. The set of bands that are connected to the $7$th band
can not be assigned to any (quasi-)BR based on the available information
and are grouped in the last entry (to be assigned).}
\end{figure}

For the topological exploration, the Fourier coefficients $A(\mathbf{k})$
of the random potentials are extracted from the distribution in Eq.
(\ref{eq:FourierCompDist}) with $\alpha=1/2$ and $C=2$. For each
potential we have calculated the symmetry fingerprint for the first
two sets of bands and used it to identify the set with one of the
eight EBRs, a composite EBR, or a topological set of bands if the
first two options were excluded. All topological sets of bands discovered
corresponded to only two different symmetry fingerprints. To determine
the fingerprints two ingredients are required (both represented in
the symmetry tables in Fig 3(a) of the main text): (i) the irreps
for each band at each high symmetry point. This information is provided
directly from the NN; (ii) the connectivity of the bands, i.e. which
pairs of bands are connected somewhere in the BZ. This information
is not provided directly by the NN but has to be inferred by looking
at $\pi$-defects in the Berry flux on a fine ${\bf k}$-grid, more
on this later, or by using our conjectured compatibility relation. 

To calculate the Berry flux, we divide the BZ in small rectangular
plaquettes $\mathbf{j}$. The Berry flux $\Phi_{\mathbf{j}}$ is just
the Berry phase acquired while encircling each plaquette. It can be
easily calculated numerically using the formula 
\begin{equation}
\Phi_{\mathbf{j}}=i\ln\left(\frac{o_{\mathbf{j},\mathbf{e}_{x}}^{(l)}o_{\mathbf{j}+\mathbf{e}_{x},\mathbf{e}_{y}}^{(l)}o_{\mathbf{j}+\mathbf{e}_{x}+\mathbf{e}_{y},-\mathbf{e}_{x}}^{(l)}o_{\mathbf{j}+\mathbf{e}_{y},-\mathbf{e}_{y}}^{(l)}}{|o_{\mathbf{j},\mathbf{e}_{x}}^{(l)}o_{\mathbf{j}+\mathbf{e}_{x},\mathbf{e}_{y}}^{(l)}o_{\mathbf{j}+\mathbf{e}_{x}+\mathbf{e}_{y},-\mathbf{e}_{x}}^{(l)}o_{\mathbf{j}+\mathbf{e}_{y},-\mathbf{e}_{y}}^{(l)}|}\right),\label{eq:Berry_flux}
\end{equation}
where $o_{\mathbf{j},\Delta\mathbf{j}}^{(l)}=\langle\mathbf{k}_{\mathbf{j}}^{(l)}|\mathbf{k}_{\mathbf{j}+\Delta\mathbf{j}}^{(l)}\rangle$,
$\mathbf{e}_{x}=(1,0)$, and $\mathbf{e}_{y}=(0,1)$. This method
allows to find connections efficiently because $|\Phi_{\mathbf{j}}|\approx\pi$
whenever a Dirac cone is inside the plaquette and the plaquette is
so small that the band dispersion can be approximated as linear (this
can be proven by approximating the Hamiltonian with a Dirac Hamiltonian).
The requirement that the band dispersion should be linear inside a
plaquette containing a Dirac cone determines how fine the grid should
be to obtain reliable results. Whether this requirement is satisfied
for a given grid depends on the specific potential. For this reason,
even though a coarse grid would be already enough to obtain reliable
results for the majority of the potentials, a very fine grid is necessary
to get high accuracy statistical results (much finer than the grid
used for training). Moreover, even for a fine grid the method might
fail in a handful of statistically irrelevant cases. For this reason,
it would be difficult to obtain the results shown in the main text
without relying on the speed of our NN.

A much faster method to calculate the connectivity is to infer it
from the compatibility relations, including also the additional parity
compatibility relation. We have compared the results obtained with
this method to the one obtained with the Berry flux method using a
fine grid containing $\sim10000$ plaquettes, cf Fig. \ref{fig:add_results_statistics}.
We found a disagreement in only less than $0.1\%$ $(1\%)$ of the
cases for the first (second) group of bands. This indicates that the
compatibility relations (when also the additional parity comaptibility
relation is taken into account) are sufficient to identify connected
bands for the overwhealming majority of the samples.
\begin{figure}
\includegraphics[width=1\columnwidth]{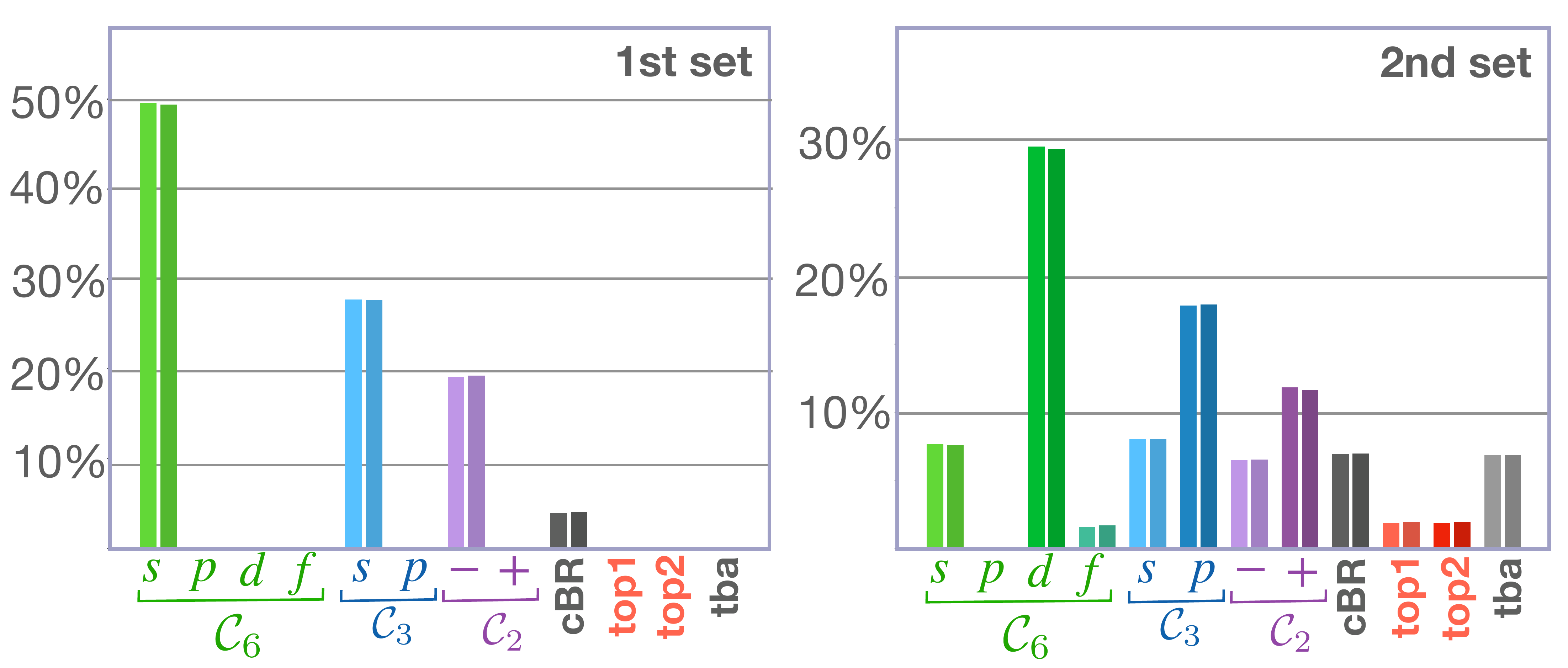}

\caption{\label{fig:comparison-NN-SE}Comparison of the (quasi-)BR distribution
calculated using the NN and the Schrödinger equation. The probability
bars for a given (quasi-)BR but different methods are shown side by
side. The results obtained using the Schrödinger equation are plotted
in slightly darker colors.}
\end{figure}

The speed and reliability of the compatibility method offer the possibility
to validate the results obtained using the NN with results obtained
directly solving the Schrödinger equation, cf Fig. \ref{fig:comparison-NN-SE}.

\begin{figure*}
\includegraphics[width=0.8\textwidth]{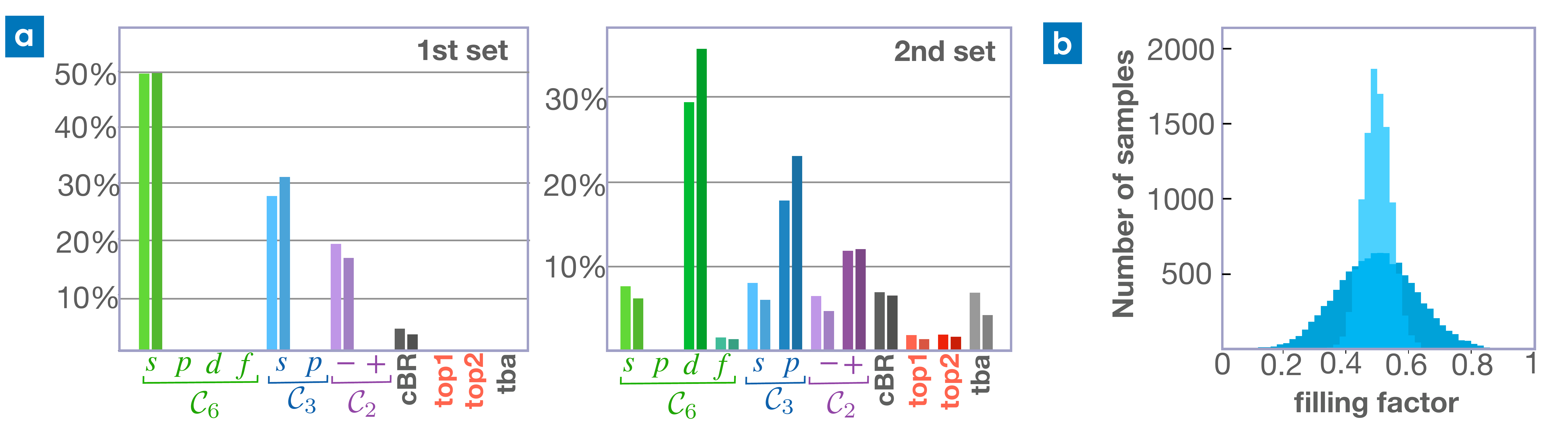}

\caption{\label{fig:add_results_statistics-1}(a) Comparison of the (quasi)-BR
distributions for two different distributions of potentials. The probability
bars for a given (quasi-)BR but different distributions are shown
side by side. (b) Number of samples as a function of the filling factor
(portion of the unit cell where $V(\mathbf{x})=V_{{\rm max}}$) for
the two distributions of interest. In all panels the results referring
to the training distribution are plotted in slightly darker colors.}
\end{figure*}

In order to give an idea how the distribution of (quasi-)BRs depends
on the underlying potential distribution we have calculated the statistics
also for the training distribution. A comparison between the two statistics
is shown in Fig. \ref{fig:add_results_statistics-1}

\section{\textcolor{black}{Supplemental Information for the 3D non-symmorphic
case study}}

\textcolor{black}{\label{sec:Details-of-the-3D}}

\subsection{\textcolor{black}{Symmetry-enhanced TB model }}

\textcolor{black}{For our two symmorphic examples we have constructed
symmetry-enhanced TB models that are defined on the Bravais lattice
of the underlying space group. As a consequence, the orbitals are
representations of its point group. This is not possible for the space
group $p4_{2}22$ (or any other non-symmorphic group) as it is well
known that TB models with non-symmorphic space groups should be defined
on a lattice with a basis. Instead, we generalize our approach by
considering as a lattice a single so-called chrystallographic orbit
(the set of sites obtained by applying all space-group transformations
to a single site). In this case, the number of sublattices is the
so-called multiplicity of the underlying Wyckoff position while the
orbitals are representations of the corresponding site-symmetry group
(the group of transformations that leaves a site invariant), which
is (isomorphic to) a subgroup of the point group to which the space
group belongs. For non-symmorphic groups, the multiplicity is larger
than one for all Wyckoff positions leading to a lattice with a basis.
(This in contrast to symmorphic groups where there is at least one
position with multiplicity one.) }

\textcolor{black}{}
\begin{table}
\centering{}\textcolor{black}{}%
\begin{tabular}{|c|c|ccccc|}
\hline 
\multicolumn{2}{|c|}{\textcolor{black}{Orb.}} & \textcolor{black}{$1\!\!1$} & \textcolor{black}{$(R_{\pi,z},\boldsymbol{a}_{3})$} & \textcolor{black}{$R_{\pi,y}$} & \textcolor{black}{$(R_{\pi,x+y},-\boldsymbol{a}_{3}/2)$} & \textcolor{black}{$(R_{\pi/2,z},\boldsymbol{a}_{3}/2)$}\tabularnewline
\hline 
\multirow{2}{*}{\textcolor{black}{$A$}} & \textcolor{black}{$+$} & \textcolor{black}{$1$} & \textcolor{black}{$1$} & \textcolor{black}{$1$} & \textcolor{black}{$1$} & \textcolor{black}{$1$}\tabularnewline
\cline{2-7} \cline{3-7} \cline{4-7} \cline{5-7} \cline{6-7} \cline{7-7} 
 & \textcolor{black}{$-$} & \textcolor{black}{$1$} & \textcolor{black}{$1$} & \textcolor{black}{$1$} & \textcolor{black}{$-1$} & \textcolor{black}{$-1$}\tabularnewline
\hline 
\multirow{2}{*}{\textcolor{black}{$B_{1}$}} & \textcolor{black}{$+$} & \textcolor{black}{$1$} & \textcolor{black}{$1$} & \textcolor{black}{$-1$} & \textcolor{black}{$-1$} & \textcolor{black}{$1$}\tabularnewline
\cline{2-7} \cline{3-7} \cline{4-7} \cline{5-7} \cline{6-7} \cline{7-7} 
 & \textcolor{black}{$-$} & \textcolor{black}{$1$} & \textcolor{black}{$1$} & \textcolor{black}{$-1$} & \textcolor{black}{$1$} & \textcolor{black}{$-1$}\tabularnewline
\hline 
\multicolumn{2}{|c|}{\textcolor{black}{$B_{2}$}} & \multirow{2}{*}{\textcolor{black}{$2$}} & \multirow{2}{*}{\textcolor{black}{$-2$}} & \multirow{2}{*}{\textcolor{black}{$0$}} & \multirow{2}{*}{\textcolor{black}{$0$}} & \multirow{2}{*}{\textcolor{black}{$0$}}\tabularnewline
\cline{1-2} \cline{2-2} 
\multicolumn{2}{|c|}{\textcolor{black}{$B_{3}$}} &  &  &  &  & \tabularnewline
\hline 
\end{tabular}\textcolor{black}{\caption{Character table of the irreps of the little group $G_{\mathbf{k}}$
for $\mathbf{k}=\Gamma,M$. For the first column, each row corresponds
to an EBR as identified by the orbital-type (i.e. the corresponding
irreps of ${\cal D}_{2}$) on sublattice $s=0$. For the remaining
columns, each row corresponds to a different irrep of $G_{\mathbf{k}}$.
For the cases in which the same EBR give rise to two one-dimensional
irreps of $G_{\mathbf{k}}$, the wavefunction is either a symmetric
(indicated with a $+$ sign in the table) or an anti-symmetric ($-$
in the table) superposition of sub-lattice polarized plane waves.
\label{tab:Character-table-for-gamma-and-M}}
}
\end{table}

\textcolor{black}{}
\begin{table}
\begin{centering}
\textcolor{black}{}%
\begin{tabular}{|c|c|ccccc|}
\hline 
\multicolumn{2}{|c|}{\textcolor{black}{Orb.}} & \textcolor{black}{$1\!\!1$} & \textcolor{black}{$(R_{\pi,z},\boldsymbol{a}_{3})$} & \textcolor{black}{$R_{\pi,y}$} & \textcolor{black}{$(R_{\pi,x+y},-\boldsymbol{a}_{3}/2)$} & \textcolor{black}{$(R_{\pi/2,z},\boldsymbol{a}_{3}/2)$}\tabularnewline
\hline 
\multicolumn{2}{|c|}{\textcolor{black}{$A$}} & \multirow{2}{*}{\textcolor{black}{$2$}} & \multirow{2}{*}{\textcolor{black}{$-2$}} & \multirow{2}{*}{\textcolor{black}{$0$}} & \multirow{2}{*}{\textcolor{black}{$0$}} & \multirow{2}{*}{\textcolor{black}{$0$}}\tabularnewline
\cline{1-2} \cline{2-2} 
\multicolumn{2}{|c|}{\textcolor{black}{$B_{1}$}} &  &  &  &  & \tabularnewline
\hline 
\multirow{2}{*}{\textcolor{black}{$B_{2}$}} & \textcolor{black}{$+$} & \textcolor{black}{$1$} & \textcolor{black}{$1$} & \textcolor{black}{$1$} & \textcolor{black}{$1$} & \textcolor{black}{$1$}\tabularnewline
\cline{2-7} \cline{3-7} \cline{4-7} \cline{5-7} \cline{6-7} \cline{7-7} 
 & \textcolor{black}{$-$} & \textcolor{black}{$1$} & \textcolor{black}{$1$} & \textcolor{black}{$1$} & \textcolor{black}{$-1$} & \textcolor{black}{$-1$}\tabularnewline
\hline 
\multirow{2}{*}{\textcolor{black}{$B_{3}$}} & \textcolor{black}{$+$} & \textcolor{black}{$1$} & \textcolor{black}{$1$} & \textcolor{black}{$-1$} & \textcolor{black}{$-1$} & \textcolor{black}{$1$}\tabularnewline
\cline{2-7} \cline{3-7} \cline{4-7} \cline{5-7} \cline{6-7} \cline{7-7} 
 & \textcolor{black}{$-$} & \textcolor{black}{$1$} & \textcolor{black}{$1$} & \textcolor{black}{$-1$} & \textcolor{black}{$1$} & \textcolor{black}{$-1$}\tabularnewline
\hline 
\end{tabular}
\par\end{centering}
\centering{}\textcolor{black}{\caption{Characters table of the representations of the little group $G_{\mathbf{k}}$
for $\mathbf{k}=Z,A$.\label{tab:Character-table-for-Z-and-A} See
caption of Table \ref{tab:Character-table-for-gamma-and-M}.}
}
\end{table}

\textcolor{black}{}
\begin{table}
\begin{centering}
\textcolor{black}{}%
\begin{tabular}{|c|c|cccc|c|}
\hline 
\multicolumn{2}{|c|}{\textcolor{black}{Orb.}} & \textcolor{black}{$1\!\!1$} & \textcolor{black}{$R_{\pi,z}$} & \textcolor{black}{$R_{\pi,y}$} & \textcolor{black}{$R_{\pi,x}$} & \textcolor{black}{Irrep.}\tabularnewline
\hline 
\multicolumn{2}{|c|}{\textcolor{black}{$A$}} & \textcolor{black}{$1$} & \textcolor{black}{$1$} & \textcolor{black}{$1$} & \textcolor{black}{$1$} & \textcolor{black}{$A$}\tabularnewline
\hline 
\multicolumn{2}{|c|}{\textcolor{black}{$B_{1}$}} & \textcolor{black}{$1$} & \textcolor{black}{$1$} & \textcolor{black}{$-1$} & \textcolor{black}{$-1$} & \textcolor{black}{$B_{1}$}\tabularnewline
\hline 
\multirow{2}{*}{\textcolor{black}{$B_{2}$}} & \textcolor{black}{$0$} & \textcolor{black}{$1$} & \textcolor{black}{$-1$} & \textcolor{black}{$1$} & \textcolor{black}{$-1$} & \textcolor{black}{$B_{2}$}\tabularnewline
 & \textcolor{black}{$1$} & \textcolor{black}{$1$} & \textcolor{black}{$-1$} & \textcolor{black}{$-1$} & \textcolor{black}{$1$} & \textcolor{black}{$B_{3}$}\tabularnewline
\hline 
\multirow{2}{*}{\textcolor{black}{$B_{3}$}} & \textcolor{black}{$0$} & \textcolor{black}{$1$} & \textcolor{black}{$-1$} & \textcolor{black}{$-1$} & \textcolor{black}{$1$} & \textcolor{black}{$B_{3}$}\tabularnewline
 & \textcolor{black}{$1$} & \textcolor{black}{$1$} & \textcolor{black}{$-1$} & \textcolor{black}{$1$} & \textcolor{black}{$-1$} & \textcolor{black}{$B_{2}$}\tabularnewline
\hline 
\end{tabular}
\par\end{centering}
\centering{}\textcolor{black}{\caption{Characters table of the representations of the little group $G_{\mathbf{k}}$
for $\mathbf{k}=X$. \label{tab:Character-table-of-X} Each row corresponds
to an an EBR as identified by the orbital-type on sublattice $s=0$.
For the cases in which the same EBR give rise to two one-dimensional
irreps of $G_{\mathbf{k}}$, the Bloch waves are sublattice polarized.
The sublattice ($0$ or $1$) is indicated in the first column. For
the maximal $\mathbf{k}$-point $X$ and $R$, the little group can
be decomposed as a sum of lattice translations and the point group
${\cal D}_{2}$. This allows to identify each irrep of $G_{\mathbf{k}}$
with an irrep of ${\cal D}_{2}$, indicated in the last column.}
}
\end{table}

\textcolor{black}{}
\begin{table}
\begin{centering}
\textcolor{black}{}%
\begin{tabular}{|c|c|cccc|c|}
\hline 
\multicolumn{2}{|c|}{\textcolor{black}{Orb.}} & \textcolor{black}{$1\!\!1$} & \textcolor{black}{$R_{\pi,z}$} & \textcolor{black}{$R_{\pi,y}$} & \textcolor{black}{$R_{\pi,x}$} & \textcolor{black}{Irrep.}\tabularnewline
\hline 
\multirow{2}{*}{\textcolor{black}{$A$}} & \textcolor{black}{$0$} & \textcolor{black}{$1$} & \textcolor{black}{$1$} & \textcolor{black}{$1$} & \textcolor{black}{$1$} & \textcolor{black}{$A$}\tabularnewline
\cline{2-7} \cline{3-7} \cline{4-7} \cline{5-7} \cline{6-7} \cline{7-7} 
 & \textcolor{black}{$1$} & \textcolor{black}{$1$} & \textcolor{black}{$1$} & \textcolor{black}{$-1$} & \textcolor{black}{$-1$} & \textcolor{black}{$B_{1}$}\tabularnewline
\hline 
\multirow{2}{*}{\textcolor{black}{$B_{1}$}} & \textcolor{black}{$0$} & \textcolor{black}{$1$} & \textcolor{black}{$1$} & \textcolor{black}{$-1$} & \textcolor{black}{$-1$} & \textcolor{black}{$B_{1}$}\tabularnewline
\cline{2-7} \cline{3-7} \cline{4-7} \cline{5-7} \cline{6-7} \cline{7-7} 
 & \textcolor{black}{$1$} & \textcolor{black}{$1$} & \textcolor{black}{$1$} & \textcolor{black}{$1$} & \textcolor{black}{$1$} & \textcolor{black}{$A$}\tabularnewline
\hline 
\multicolumn{2}{|c|}{\textcolor{black}{$B_{2}$}} & \textcolor{black}{$1$} & \textcolor{black}{$-1$} & \textcolor{black}{$1$} & \textcolor{black}{$-1$} & \textcolor{black}{$B_{2}$}\tabularnewline
\hline 
\multicolumn{2}{|c|}{\textcolor{black}{$B_{3}$}} & \textcolor{black}{$1$} & \textcolor{black}{$-1$} & \textcolor{black}{$-1$} & \textcolor{black}{$1$} & \textcolor{black}{$B_{3}$}\tabularnewline
\hline 
\end{tabular}
\par\end{centering}
\centering{}\textcolor{black}{\caption{\label{tab:Character-table-of-R}Characters table of the representations
of the little group $G_{\mathbf{k}}$ for $\mathbf{k}=R$. See caption
of Table \ref{tab:Character-table-of-X}.}
}
\end{table}

\textcolor{black}{A natural generalization of the approach we adopted
so far is to pick (one of) the Wyckoff position(s) with smallest multiplicity.
For the space group $p4_{2}22$, we choose the Wyckoff position $2a$
(here, the number $2$ indicates the multiplicity). We construct our
TB model starting from a set of orbitals $\{|n,m\rangle\}$ localized
about the origin ($n$ is the principal quantum number and $m$ labels
the irrep of the site-symmetry group). Here, the site-symmetry group
is the point group ${\cal D}_{2}$ (two-fold rotations about the $x$,
$y$, and $z$ axes). This point group has four inequivalent irreps,
$m=A,B_{1},B_{2},B_{3}$ with the atomic $s$, $p_{z}$, $p_{y}$,
and $p_{x}$-orbitals, respectively, being representive states transforming
under these irreps. From each orbital localized about the origin,
we construct another orbital localized about the position $\boldsymbol{a}_{3}/2$
by applying a screw-rotation. Finally, we obtain a basis of Wannier
states by applying all lattice translations, 
\[
|n,\mathbf{j},s,m\rangle=(1\!\!1,\boldsymbol{a}_{\boldsymbol{j}})|n,s,m\rangle.
\]
Here, $\mathbf{j}$ indicates the unit cell and $s$the sublattice.
Thus, $\boldsymbol{a}_{\boldsymbol{j}}=j_{1}\boldsymbol{a}_{1}+j_{2}\boldsymbol{a}_{2}+j_{3}\boldsymbol{a}_{3}$,
$|n,0,m\rangle=|n,m\rangle,$ and $|n,1,m\rangle=(R_{\pi/2,z},\boldsymbol{a}_{3}/2)|n,m\rangle$.
We note that according to our definition, the states $|n,\mathbf{j},1,B_{2/3}\rangle$
are labeled according to the irreps of the state $|0,m\rangle$ not
their own. For example, the state $|\mathbf{j},0,B_{2}\rangle$ transforms
under symmetry as a $p_{y}$-orbital while the state $|\mathbf{j},1,B_{2}\rangle$,
being rotated by $90^{\circ}$, will rather transform as a $p_{x}$-orbital. }

\textcolor{black}{With the above definitions in hand we find
\begin{align*}
H_{n,m;n',m'} & \equiv\langle n',m'|\hat{H}|n,m\rangle=\langle n',\mathbf{j},s,m'|\hat{H}|n',\mathbf{j},s,m'\rangle\\
 & =\omega_{n,m}\delta_{n,n'}\delta_{mm'}.
\end{align*}
Thus, we have $\tilde{n}_{A}^{({\rm TB})}+\tilde{n}_{B_{1}}^{({\rm TB})}+\tilde{n}_{B_{2}}^{({\rm TB})}+\tilde{n}_{B_{3}}^{({\rm TB})}$
independent onsite energies where $\tilde{n}_{j=A,B_{1},B_{2},B_{3}}^{({\rm TB})}$
are the number of orbitals of each type. Here we have assumed without
loss of generality that the principal basis diagonalize the Hamiltonian
projected onto the Wannier states localized about the origin. Next
we define the matrix containing the hopping amplitudes for nearest
neighbor vertical transitions within the same unit cell
\begin{equation}
J_{n'm';nm}^{(z)}\equiv\langle n',1,m'|\hat{H}|n,0,m\rangle.\label{eq:def_J_z}
\end{equation}
We note that $J_{nm;n'm'}^{(z)}$ can be chosen real because the orbitals
$|n,s,m\rangle$ are invariant under time-reversal symmetry. Moreover,
the symmetry under two-fold rotations about the $z$-axis gives rise
to a selection rule forbidding vertical hopping transitions between
states of opposite parity. Thus, for example an excitation can hop
vertically from a $A$-orbital to another $A$-orbital or a $B_{1}$-orbital
(as both $A$ and $B_{1}$ orbitals are even under two-fold rotations
about the $z$-axis) but not to a $B_{2}$ or a $B_{3}$-orbital.
Thus, the matrix $J_{n'm';nm}^{(z)}$ is block-diagonal with one block
spanned by the $A$ and $B_{1}$ orbitals and the other Block spanned
by the $B_{2}$ and $B_{3}$ orbitals. By applying a screw rotation
followed by a $180$ rotation about the $x$-axis we find
\[
J_{n'm';nm}^{(z)}=m'_{z}m_{x}'m_{y}J_{nm;n'm'}^{(z)}.
\]
Here, $m_{x/y/z}$ is the parity of the orbitals $|n,m\rangle$ under
$x/y/z$ twofold rotations. Thus, $m_{z/y/x}=1$ for $m=A,B_{1/2/3}$
and $-1$ otherwise. We can conclude, the hopping matrix $J_{n'm';nm}^{(z)}$
has $(\tilde{n}_{A}^{({\rm TB})}+\tilde{n}_{B_{1}}^{({\rm TB})}+1)(\tilde{n}_{A}^{({\rm TB})}+\tilde{n}_{B_{1}}^{({\rm TB})})/2+(\tilde{n}_{B_{2}}^{({\rm TB})}+\tilde{n}_{B_{3}}^{({\rm TB})}+1)(\tilde{n}_{B_{2}}^{({\rm TB})}+\tilde{n}_{B_{3}}^{({\rm TB})})/2$
independent matrix elements. Applying the symmetry under two-fold
transitions about the $y$-axis one finds the hopping amplitudes for
vertical NN transitions between sites in different unit cells 
\[
\langle n,\boldsymbol{a}_{z},0,m|\hat{H}|n',1,m'\rangle=J_{n'm';nm}^{(z)}m'_{x}m_{y}.
\]
Thus, all nearest-neighbor vertical hopping transitions are encoded
in the matrix $J_{n'm';nm}^{(z)}$ .}

\textcolor{black}{Next we define the horizontal hopping rates on sublattice
$s=0$ 
\begin{equation}
J_{n'm';nm}^{(x/y,0)}\equiv\langle n',\mathbf{e}_{x/y},0,m'|\hat{H}|n,0,m\rangle.\label{eq:def_J_x}
\end{equation}
Analogously to what discussed above for the vertical hopping rates,
the matrix $J_{n'm';nm}^{(x,0)}$ ($J_{n'm';nm}^{(y,0)}$) is block
diagonal, this time with one block spanned by the $A$ and the $B_{3}$
($B_{2}$) orbitals and the other by the $B_{1}$ and $B_{2}$ ($B_{3}$)
orbitals. As their vertical counterparts, the blocks of $J_{n'm';nm}^{(x/y)}$
are constrained by the symmetry
\begin{align*}
J_{n'm;nm}^{(x/y)} & =J_{nm;n'm}^{(x/y)}\\
J_{n'm';nm}^{(x/y)} & =-J_{nm;n'm'}^{(x/y)}\quad{\rm for}\quad m\neq m'.
\end{align*}
Thus, the hopping matrices $J_{n'm';nm}^{(x/y)}$ are defined by $(\tilde{n}_{A}^{({\rm TB})}+\tilde{n}_{B_{3/2}}^{({\rm TB})}+1)(\tilde{n}_{A}^{({\rm TB})}+\tilde{n}_{B_{3/2}}^{({\rm TB})})/2+(\tilde{n}_{B_{1}}^{({\rm TB})}+\tilde{n}_{B_{2/3}}^{({\rm TB})}+1)(\tilde{n}_{B_{1}}^{({\rm TB})}+\tilde{n}_{B_{2/3}}^{({\rm TB})})/2$
independent parameters. By using a screw rotation we can also obtain
the horizontal hopping matrices $J_{n'm';nm}^{(x/y,1)}$ on sublattice
$s=1$, 
\begin{align}
 & J_{n'm';nm}^{(y,1)}\equiv\langle n',\mathbf{e}_{y},1,m'|\hat{H}|n,1,m\rangle=J_{n'm';nm}^{(x,0)}.\label{eq:def_J_x-1-1}\\
 & J_{n'm';nm}^{(x,1)}\equiv\langle n',\mathbf{e}_{x},1,m'|\hat{H}|n,1,m\rangle=J_{nm;n'm'}^{(y,0)}.
\end{align}
}

\textcolor{black}{After setting the appropriate constraints to the
hopping matrices, our symmetry-enhanced TB Hamiltonian will have automatically
a block diagonal form at the maximal $\mathbf{k}$-points, with each
block corresponding to an irrep of the so-called little group $G_{\mathbf{k}}$
(the group of symmetries $g$ that leave $\mathbf{k}$ invariant modolus
a vector of the reciprocal lattice). The same property, obviously,
will hold for the Hamiltonian of the $3$D Schrödinger equation. A
crucial step in implementing our method is to require (via the appropriate
cost function contribution) that matching blocks (corresponding to
the same irrep) of both Hamiltonians have the same spectrum. Thus,
we need to identify the underlying irrep for each such block. For
this purpose, we construct the charachters Tables \ref{tab:Character-table-for-gamma-and-M},\ref{tab:Character-table-for-Z-and-A},
\ref{tab:Character-table-of-X}, and \ref{tab:Character-table-of-R}.
The first column in every table is the irrep (orbital-type) of an
orbital at the origin ($A$, $B_{1}$, $B_{2}$ or $B_{3}$). This
orbital is then used to construct the corresponding EBR. At each maximal
$\mathbf{k}$-point ($\mathbf{k}=\Gamma,X,M,Z,R,$ or $A$), such
an EBR will give rise to a representation of the little group $G_{\mathbf{k}}$
spanned by the sublattice plane-waves
\[
|n,\mathbf{k},s,m\rangle=\sum_{\mathbf{j}}e^{i\mathbf{k}\cdot\boldsymbol{a}_{\mathbf{j}}}|n,\mathbf{j},s,m\rangle,\quad s=0,1.
\]
This is either a two-dimensional irrep. of $G_{\mathbf{k}}$ or can
be decomposed in two one-dimensional irreps. In the latter case, we
also give the underlying superposition of sublattice plane-waves for
each irrep (also in the first column). As usual, it is possible to
uniquely identify any irrep of the little group by listing the characters
(traces) $\chi(\rho(g))$ of the matrix representative $\rho(g)$
for an appropriate finite set of transformations $g$. This set is
finite because it is enough to consider a single transformation for
each infinite set of transformations that differ by a lattice translation
(as the traces for different elements will differ only by a phase
that does not depend on the irrep). In addition, it is enough to consider
a single transformation for each conjugacy class (because the trace
is the same for all elements of the same conjugacy class).}

\textcolor{black}{We use the Tables \ref{tab:Character-table-for-gamma-and-M},\ref{tab:Character-table-for-Z-and-A},
\ref{tab:Character-table-of-X}, and \ref{tab:Character-table-of-R}
for three purposes: i) We identify the subset of Wannier orbitals
(or superposition thereof) that span the same block of the symmetry-enhanced
TB model at a specific high-symmetry point. For example, according
to Table \ref{tab:Character-table-for-gamma-and-M} all $B_{2}$ and
$B_{3}$ orbitals give rise to the same irrep at the $\Gamma$ and
$M$ point and, thus, will span a single block there. (ii) We assign
each exact solution of the $3$D Schrödinger equation (at a maximal
$\mathbf{k}$-point) to the correct irreps by checking its behavior
under the transformations listed in the relevant table. (iii) We calculate
the symmetry fingerprints for the EBRs generated by each orbital type.
These are then used to calculate lower bounds for $\tilde{n}_{l}^{({\rm TB})}$,
see discussion in Appendix \ref{sec:Symmetry-enhanced-tight-binding}.
As usual the lower bound depends on the number of target bands. For
four bands, the lower bounds are $\tilde{n}_{A}^{({\rm TB})}=3$,
$\tilde{n}_{B_{1}}^{({\rm TB})}=2$, $\tilde{n}_{B_{2}}^{({\rm TB})}=1$,
$\tilde{n}_{B_{3}}^{({\rm TB})}=2$. Empirically we have found that
it is advantageous to use a larger Hilbert space with $\tilde{n}_{A}^{({\rm TB})}=5$,
$\tilde{n}_{B_{1}}^{({\rm TB})}=4$, $\tilde{n}_{B_{2}}^{({\rm TB})}=3$,
$\tilde{n}_{B_{3}}^{({\rm TB})}=4$.}

\subsection{\textcolor{black}{NN layout }}

\begin{figure}
\textcolor{black}{\includegraphics[width=1\columnwidth]{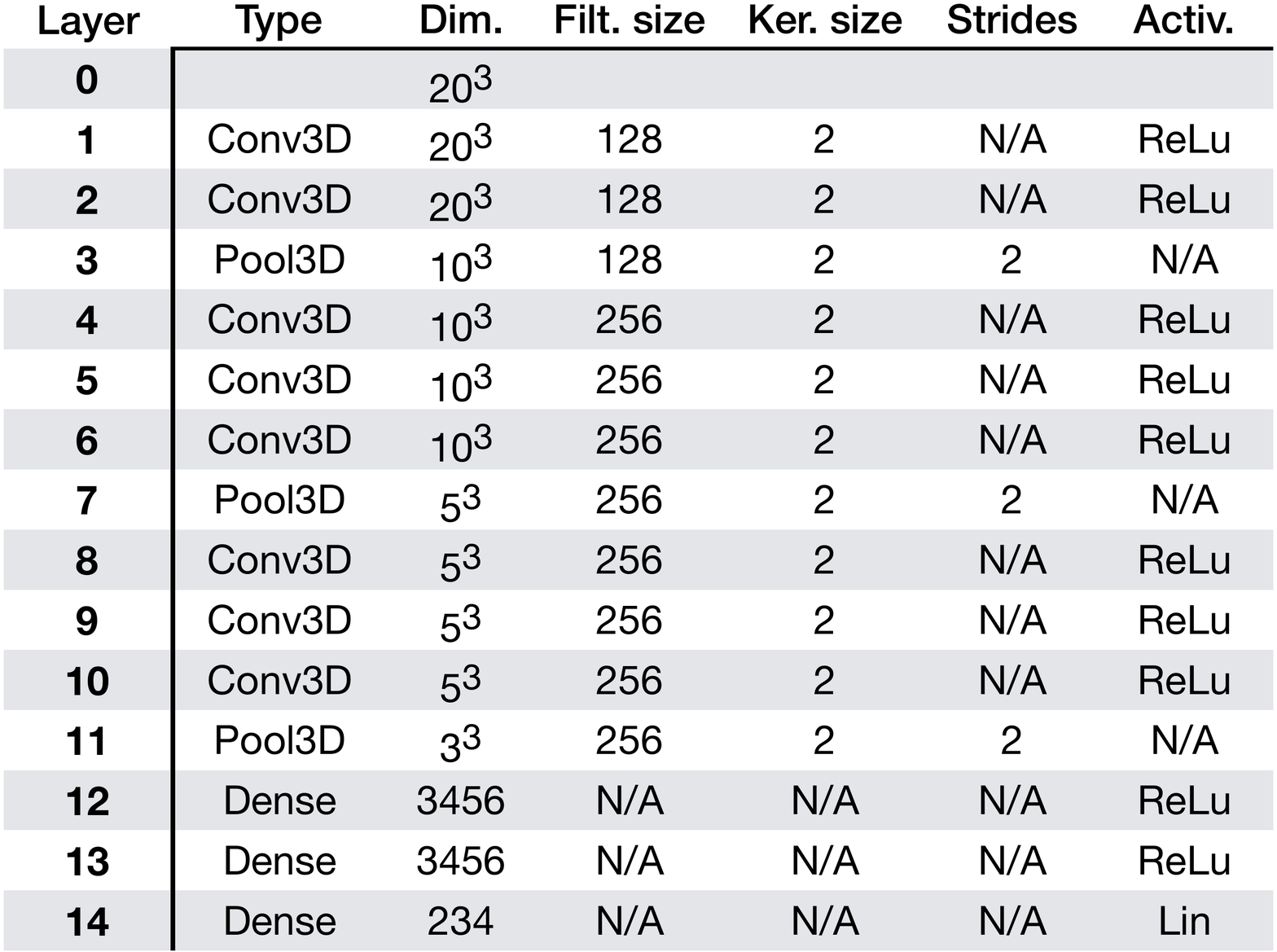}\caption{\label{fig:NNlayout-1}Sketch of the NN layout for the 3D non-symmorphic
case study. }
}
\end{figure}

\textcolor{black}{The NN layout for the $3$D case study is sketched
in Fig. \ref{fig:NNlayout-1}. In this case, we have used }\lstinline!conv3D!\textcolor{black}{{}
layers with, RELU activation, Kernel size $(2,2,2)$ using the option
}\lstinline!padding=same!\textcolor{black}{. With the aim of progressively
reducing the image size, the convolutional layers are alternated with
}\lstinline!max_pooling3D!\textcolor{black}{{} layers with Pool size
$(2,2,2)$ and }\lstinline!stride=2!\textcolor{black}{. As usual
the convolutional and pooling layers are followed by a series of dense
layers. The }\lstinline!dropout(0.15)!\textcolor{black}{{} is applied
between each pair of subsequent dense layers (not shown in the sketch).}

\textcolor{black}{}

\section{\textcolor{black}{Supplemental Information for the two-component
Topological Metamaterial case study\label{sec:Details-of-the-Spin_Hall}}}

\subsection{\textcolor{black}{Details of the microscopic model}}

\textcolor{black}{As discussed in the main text we have modeled one
spin sector of our two-component metamaterial using 'one-half' of
the Bernevig-Hughes-Zhang model with a site dependent mass $M_{\mathbf{j}}$,
\begin{equation}
H=\sum_{\mathbf{j}}(M_{\mathbf{j}}+2J)\hat{\sigma}_{z}|\mathbf{j}\rangle\langle\mathbf{j}|-\frac{J}{2}\sum_{\langle\mathbf{l},\mathbf{j}\rangle}(\hat{\sigma}_{z}-i\hat{\sigma}_{d(\mathbf{l},\mathbf{j})})|\mathbf{l}\rangle\langle\mathbf{j}|+h.c.\label{eq:BHZ_Hamiltonian}
\end{equation}
Here, $M_{\mathbf{j}}=\pm m$ where the multiindex $\mathbf{j}=(j_{x,}j_{y})$
parametrizes the sites of a $N\times N$ square grid, $\langle\mathbf{l},\mathbf{j}\rangle$
indicates the sum over nearest-neighbors, $d(\mathbf{l},\mathbf{j})=x,y$
is the hopping direction, and $J$ is the hopping rate. In our simulation,
we choose $N=70$ and $m/J=10/N\approx0.14$ with $J>0$. For this
sign of $J$, any positive mass $M>0$ corresponds to the trivial
phase, ${\cal C}=0$. On the other hand, for the topological region
the mass $M=-m$ falls in the interval $-2<M/J<0$ where ${\cal C}=1$
\citep{bernevig_quantum_2006}. We note that the Hamiltonian (\ref{eq:BHZ_Hamiltonian})
applies not only to a scenario in which the spin is conserved but
also to a more general case in which the spin is not necessarily conserved
but the out-of-plane mirror transformation $\hat{M}_{z}$ is still
a symmetry. Thus, also our results apply to this more general scenario.
In this framework the 'half' HBZ Hamiltonian as well as the corresponding
Chern numbers refer to a Mirror symmetry sector (one of the two possible
eigenvalues of the Mirror symmetry, $M_{z}=i$ or $M_{z}=-i$ ). }

\textcolor{black}{Taking the large wavelength limit of the 'half'
BHZ Hamiltonian Eq. (\ref{eq:BHZ_Hamiltonian}), one arrives to the
Dirac Hamiltonian Eq. (\ref{eq:Dirac_Ham}) with $v=Ja/N$. It is
of fundamental interest to investigate the large wavelength limit
because it is of higher generality going beyond our specific microscopic
model. In particular, the physics becomes independent of $N$ (once
$v/m$ is held fixed). Our particle-hole conserving Dirac Hamiltonian
would also emerge as a large wavelength description of a more general
form of the BHZ Hamiltonian that does not have this symmetry (as in
the original formulation of this model). More in general, it will
describe any situation where the two component materials differ in
Chern number by one unit and their valence and conduction bands have
minimal splitting at the $\Gamma$ point where they support gapped
Dirac cones, irrespective of the underlying microscopic lattice. To
make sure that our results are not model specific and really apply
to the most general setting described above, we have checked that
they have converged to the limit $N\to\infty$ comparing simulations
with equal $v/M$ but different $N$.}

\subsection{\textcolor{black}{Symmetry-enhanced TB model}}

\textcolor{black}{We construct our symmetry enhanced TB model using
orbitals localized about the Wyckoff position $1a$ (one of the two
fourfold rotocenters). The orbitals will then be irreps of the point
group ${\cal C}_{4}$. In this case, the relevant irreps are not time-reversal
symmetric and, thus, are simply parametrized by the quasi-angular
momentum $m$, $m=0,\pm1,2\:{\rm mod\,4}$, or $s$, $p_{\pm}$ and
$d$. The particle-hole symmetry sets the additional constraint that
pairs of orbitals with quasi-angular momentum $m\,{\rm mod\,4}$ and
$1-m\,{\rm mod}\,4$, respectively, have opposite energy. Thus, there
are only two different types of orbital pairs (or equivalently EBRs):
for one EBR the quasi-angular momenta of the two particle-hole-connected
orbitals are $m=0$ and $m=-1$, for the other $m=2$ and $m=1$.
The constraints described above lead to a symmetry-enhanced TB Hamiltonian
in the form}

\textcolor{black}{
\[
\hat{H}_{\mathbf{k};n,m;n'm'}(\mathbf{k)}/\hbar=\delta_{m,m}\delta_{n,n'}\omega_{n,m}+J_{n,m;n',m'}f_{m-m'}(\mathbf{k}).
\]
Because of the ${\cal C}_{4}$ symmetry we find
\begin{align*}
 & f_{0}(\mathbf{k})=\cos(k_{x}a)+\cos(k_{y}a),\\
 & f_{1}(\mathbf{k})=-f_{-1}^{*}(\mathbf{k})=-i[\sin(k_{x}a)+e^{-i\pi/2}\sin(k_{y}a)],\\
 & f_{2}(\mathbf{k})=\cos(k_{x}a)-\cos(k_{y}a).
\end{align*}
where $J_{n,m;n',m'}$ are the hoppings in the righward direction.
Note that from the inversion symmetry it follows that 
\[
J_{n,m;n',m'}=\pm J_{n',m';n,m}^{*},
\]
where the positive (negative) sign applies if $\Delta m=m-m'$ is
even (odd). {[}This constraint ensues that the Hamiltonian is Hermitian.{]}
Because of the particle hole symmetry we have the additional constraints
\[
\omega_{n,m}=-\omega_{n,-1-m},\quad J_{n,m;n,m'}=-J_{n,-1-m;n',-1-m'}^{*}.
\]
This means that the independent coefficients can be chosen to be $\omega_{n,m}$
(real) for $m=0,2$, $J_{n,m;n',m'}$ (complex) for:
\begin{align*}
 & m=0,\quad m'=0,\quad{\rm for}\quad n>n',\\
 & m=2,\quad m'=2,\quad{\rm for}\quad n>n',\\
 & m=0,\quad m'=-1,\quad{\rm for}\quad n\ge n'\\
 & m=1,\quad m'=2,\quad{\rm for}\quad n\ge n'\\
 & m=0,\quad m'=2\\
 & m=0,\quad m'=1,
\end{align*}
and $J_{n,m;n,m}$ (real) for $m=0,2$. The number of independent
parameters is thus $2(\tilde{n}_{s}^{({\rm TB)}}+\tilde{n}_{d}^{({\rm TB)}})(1+\tilde{n}_{s}^{({\rm TB)}}+\tilde{n}_{d}^{({\rm TB)}})$
where ($\tilde{n}_{d}^{({\rm TB)}}$) $\tilde{n}_{s}^{({\rm TB)}}$
is the number of ($d$-) $s$-orbitals. For training on the $8$ central
bands we have used $\tilde{n}_{s}^{({\rm TB)}}=9$ and $\tilde{n}_{d}^{({\rm TB)}}=8$.}

\textcolor{black}{In this case study, it is straightforward to match
different blocks of the TB model Hamiltonian with the corresponding
blocks of the Dirac equation by checking for the quasi-angular momentum
(parity) at the $\Gamma$ and $M$ points ($X$ point). However, there
remain an oustanding challenge: since the two models have a different
numbers of bands and we are trying to predict the band structure in
the middle of the spectrum, it is still not obvious which energy level
should correspond to which. For the global cost function Eq. (\ref{eq:cost}),
this ambiguity is readily eliminated by the paricle-hole-symmetry:
by symmetry there is always an equal number of positive and negative
energy states and one can simply match the first positive bands of
the symmetry-enhanced TB model to the corresponding levels of the
Dirac equation. At a maximal $\mathbf{k}$-point, however, this symmetry
does not apply separately to each block. Since the particle-hole symmetry
maps onto each other states of different symmetry, e.g. $s$ and $p_{-}$,
there is no guarantee that the number of positive and negative energy
states of a given symmetry, e.g. $p_{-}$, are the same. In this case,
the matching requires some additional assumption. We assume that a
band touching between the higher-energy negative band and the lowest-energy
positive band (leading to a band inversion) can occur only at the
$\Gamma$ point between a $s$ and a $p_{-}$ Bloch wave. In other
words, we assume that for each irrep the number of positive and negative
bands is the same with a single exception: whenever the $p_{-}$ Bloch
wave is the lowest positive energy state at the $\Gamma$ point, we
allow for one additional positive (negative) energy $p_{-}$-state
($s$-state). We have not proven this assumption but we believe that
the nearly perfect Chern number predictions by post-selected NNs is
a good indication of its validity. }

\textcolor{black}{In the main text, we have compared the prediction
of our NN to microscopic simulations for the band Chern number. This
can readily be calculated as the sum of the Berry fluxes, cf (\ref{eq:Berry_flux}),
over all quasi-momentum-plaquettes of a fine grid (here a $62\times62$
grid) covering the whole BZ \citet{fukui_chern_2005}. Alternatively,
one might have considered the band gap Chern number, this is the sum
of the band Chern numbers of all bands below a certain band gap. We
focused on the band Chern number mainly for two reasons discussed
below. The first reason is of fundamental nature: The band Chern numbers
of the central bands are well defined within the Dirac Hamiltonian
large-wavelength description. On the other hand, the band gap Chern
number goes beyond the long wave length limit in that it involves
also bands with large negative energy outside of the bandwidth where
the large wavelenght limit is expected to apply. The second motivation
to focus on the band Chern numbers is of practical nature: the band
Chern number are easily accessible by calculating the band structure
and eigenvectors in the central region of the spectrum using the Lanczos
algorithm. On the other hand, the calculation of the band gap Chern
numbers would be numerically very expensive because of the large number
of bands of our microscopic model (we have $2\times N^{2}$ bands
with $N=70$ in our simulations).}

\textcolor{black}{While, for the reasons discussed above, we have
focused on the band Chern number so far, we wish to address at least
briefly the question whether the band gap Chern numbers predicted
by our symmetry-enhanced TB model do coincide with those of the microscopic
model. This is an important question because the band gap Chern numbers
are relevant for the bulk-boundary correspondence and, thus, determines
the behavior of the edge states in a system with boundary. We note
that the assumption that a band touching between the lowest-energy
positive band and the highest-energy negative band occurs only between
an $s$ and a $p_{-}$ band at the $\Gamma$ point (used while matching
the bands of our symmetry-enhanced TB model to the bands of the microscopic
model), represents a stringent constraint to the Chern number ${\cal C}_{-/+}$
for the band gap separating the positive bands from the negative bands
(or, equivalently, to the sum of the Chern numbers over all negative-energy
bands). If this assumption holds true, there remain only two scenarios:
(i) ${\cal C}_{-/+}=0$ if a $s$-state is the lowest positive energy
state at the $\Gamma$ point (no band inversion); (ii) ${\cal C}_{-/+}=1$
if the lowest positive-energy eigenstate is a $p_{-}$ Bloch-wave.
This is analogous to what happens in a homogeneous bulk. In the framework
of this hypothesis, we can easily derive any band gap Chern number
(for the central bands) from the band Chern numbers of the first few
positive bands and the symmetry label of the lowest energy positive
state at the $\Gamma$ point, which are both very accurately predicted
by our NNs. }

\textcolor{black}{We have verified our hypothesis regarding the Chern
number ${\cal C}_{-/+}$ for a statistically relevant number of samples
using the bulk-boundary correspondence. This has been achieved by
deducing ${\cal C}_{-/+}$ from the slopes and position (upper or
lower edge) of the edge states in the central band gap, calculated
from strip simulations. With the aim of reducing the computational
effort we have considered a low resolution grid ($N=10$) and a narrow
strip ($10$ unit cells width). While we do not expect our strip simulations
for such a low resolution to have already converged to the continous
limit, we still expect them to reproduce at least qualitatively the
band structure and to correctly reproduce the robust topological features
of interest here. }

\subsection{\textcolor{black}{NN layout }}

\begin{figure}
\textcolor{black}{\includegraphics[width=1\columnwidth]{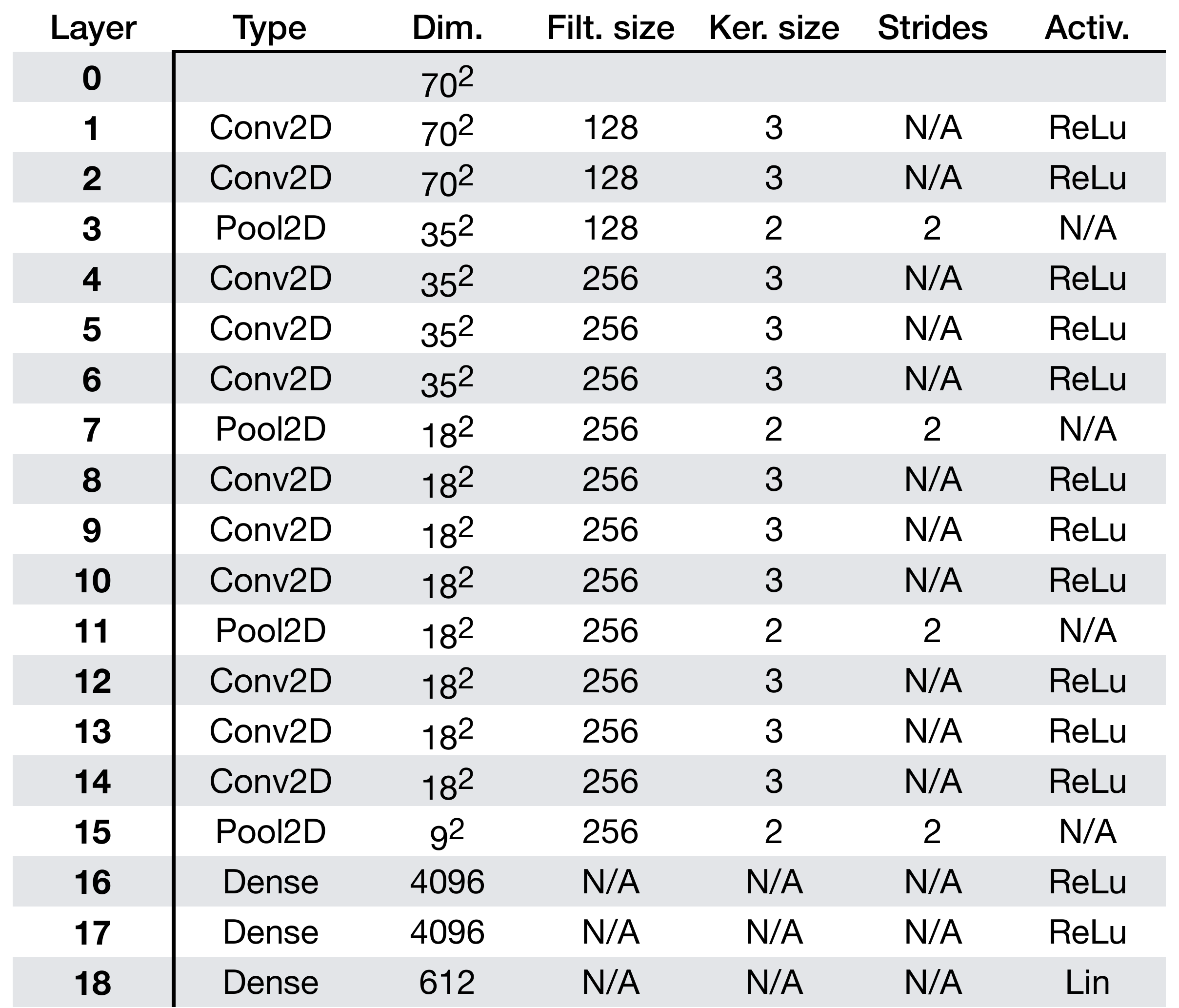}\caption{\label{fig:NNlayout-1-1}Sketch of the NN layout for the two-component
topological insulator case study.}
}
\end{figure}

\textcolor{black}{The NN layout is sketched in Fig. \ref{fig:NNlayout-1-1}.
}\lstinline!conv2D!\textcolor{black}{{} layers with RELU activation,
Kernel size $(2,2)$ using the option }\lstinline!padding=same!\textcolor{black}{{}
are alternated to }\lstinline!max_pooling2D!\textcolor{black}{{} layers
with Pool size $(2,2)$ and }\lstinline!stride=2!\textcolor{black}{.
These layers are followed by three dense layers. The }\lstinline!dropout(0.15)!\textcolor{black}{{}
is applied between each pair of subsequent dense layer (not shown
in the sketch).}

\textcolor{black}{}

\section{Details of the Optimization Method\label{sec:OptimizationApp}}

Implementing an optimization task starts by specifying some goal to
achieve for the band structure: e.g. maximizing some band gap, or
matching the predicted band structure as well as possible to a fixed
given band structure. This goal has to be expressed in terms of a
reward function (a function of the predicted band structure). Furthermore,
the geometry has to be parametrized; in our case we choose to describe
a completely general geometry via its Fourier coefficients. Afterwards,
one can do gradient ascent on the reward, with respect to the geometry,
exploiting the fact that backpropagation through the full network/tight-binding
pipeline is possible.

The illustrative example we treat in the main text is a kind of 'inverse
problem', where we want to target a given band structure (calculated
from some selected simple tight-binding model, in our case). 

\subsection{Creation of sharp potentials from smooth functions}

A difficulty in our case are the constraints on the potential, which
should assume only two discrete values, besides being C6-symmetric.
To guarantee these properties, we optimise not directly the potential
defined on a lattice, but instead the Fourier coefficients from which
the potentials can be generated. This is possible since our approach
for creating potentials, applying a sigmoid to a smooth scalar field,
is differentiable for finite (non-zero) ``temperatures'' of the
sigmoid. The step from Fourier coefficients to potential can be implemented
directly in TensorFlow.

In this way, the potentials already obey the required symmetry. However,
even though the sigmoid constrains the potential values between 0
and $V_{{\rm max}}$, it cannot reliably enforce the potential to
take only these values. In order to enforce the discreteness of the
potential, we define a new term for the cost function which is proportional
to $\sum_{\mathbf{x}}v(\mathbf{x})(v(\mathbf{x})-1)$, where $\mathbf{x}$
runs over the unit cell and $v(\mathbf{x})=V(\mathbf{x})/V_{{\rm max}}$
is the rescaled potential. The cost function for optimising the input
is hence the old cost function plus this new \emph{potential cost
function}. Again, it is important to choose the right weighting between
both contributions. Without the potential cost function, the optimised
potential might become less discrete, while a weight that is too large
leads to stagnation of training. We have observed that a good weight
for the potential cost function is around $0.0001$. In this way,
the neural network focuses on making the potentials discrete only
after it has reached already a low loss value. As optimiser we again
use Adam with a learning rate of around 0.5 and otherwise default
settings.

\subsection{Our optimization procedure}

Since it can happen that certain starting conditions may lead to stagnation,
we recommend to run several optimisation trials with different random
starting conditions. In our case, we choose a uniform distribution
for all coefficients, with 0 as mean. The suitable limits for the
uniform distribution depends on the choice of inverse temperature
$\beta$ (cf Eq. (\ref{eq:Beta_def}). In the case of $\beta=1$ we
observe good results for random amplitudes in the interval $[-0.5,0.5]$.
Since the optimisation can be performed on a GPU, we are able to optimise
many samples in parallel. For example, we can optimise for one target
band structure with 200 different starting conditions at the same
time. On our hardware (mentioned in the training section) 10 update
steps on 200 samples take roughly 3 seconds. As usual number of update
steps we recommend 400-500. Out of the 200 trials for one band structure,
we can then choose the best trials and check the results of the Schrödinger
equation on the predicted potentials.

\subsection{Details on the calculation of the loss map }

For the optimization loss map (measuring the quadratic deviation between
the optimized band structure and the target, for different target
band structures), we use parallelization to optimize for different
band structures at the same time. By this, we can quickly produce
one potential for each point in our grid. By repeating this procedure
and updating the loss map such that the loss and the corresponding
potential are replaced if the new version is better, we can reduce
the noise of the loss map over time. To assess the quality, we always
compare against the loss of the neural network predictions on validation
data (this is the square of the rms band structure deviation measured
in units of $V_{{\rm Max}}$). For the final optimization loss map,
we distinguish between ``relatively good'' results, with an optimization
loss about twice the network loss, and very good results, with an
optimization loss below the network loss (here roughly $6\cdot10^{-6}$
).

\section{Tight-binding model supporting fragile topological phases\label{sec:Fragile-Tight-binding-model}}

Here, we give more details on the TB model implemented using our optimization
method. For a detailed investigation of the model with only nearest
neighbor hopping we refer to Refs. \citep{wu_topological_2016,kariyado_topological_2017}.
There are two main reasons why this model has attracted huge attention
in the field of topological physics: i) It is the simplest toy model
that describes the band folding $p$-$d$ band inversion transition
that underlies the designs of a large number of topological photonics
and topological phononics experiments demonstrating helical edge states,
see e.g. \citep{barik_topological_2018,parappurath_direct_2018,cha_experimental_2018}.
In this case, the edge states are localized about domain walls separating
a 'trivial' region (without band inversion, $J'<J$) and a 'topological'
region (with band inversion, $J'>J$). ii) It is possible to describe
its topology in terms of mirror winding numbers. Such mirror winding
number are connected via a bulk-boundary correspondence to the edge
states at the physical boundary of systems with a selected shape (decoration)
\citep{kariyado_topological_2017}. We emphasize, however, that from
the point of view of topological quantum chemistry (which focus on
the orbitals in real space rather than the bulk boundary correspondence)
the band folding phase transition is actually a hybridization transition
and not a topological one, see \citep{de_paz_engineering_2019}. For
$J>J'$, the Wannier orbitals are formed by the hybridization of six
atoms within one unit cell and are, thus, localized about the ${\cal C}_{6}$
rotocenters. For $J>J'$, on the other hand, the Wannier orbitals
are formed by the dimerization of pairs of nearest neighbor orbitals
belonging to different unit cells and are, thus, localized about the
${\cal C}_{2}$ rotocenters. In both cases, there is a well defined
'atomic limit'.

Here, we show that by adding a next-nearest neighbor hopping modulated
in amplitude it is possible to induce a topological phase transition
where a set of isolated bands does not admit an 'atomic limit'. More
precisely, we will implement topological quasi-BRs with the same symmetry
fingerprints as those we have discovered in our topological exploration,
cf Fig. 3(d) of the main text. Thus, this finding is yet another example
of discovery that was stimulated by the rapid exploration allowed
by our NN.

The Hamiltonian for our TB model reads
\begin{align*}
H_{{\rm TB}}/\hbar= & \sum_{j}\omega_{0}\hat{a}_{j}^{\dagger}\hat{a}_{j}-\sum_{\langle j,j'\rangle}J_{j,j'}(\hat{a}_{j}^{\dagger}\hat{a}_{j'}+\hat{a}_{j'}^{\dagger}\hat{a}_{j})\\
 & -\sum_{\langle\langle j,j'\rangle\rangle}L_{j,j'}(\hat{a}_{j}^{\dagger}\hat{a}_{j'}+\hat{a}_{j'}^{\dagger}\hat{a}_{j}),
\end{align*}
where $\hat{a}_{j}$ is the annihilation operator on site $j$, $(\langle\langle j,j'\rangle\rangle)$
$\langle j,j'\rangle$ indicates the sum over (next-)nearest neighbors,
and $(L_{j,j'})$ $J_{j,j'}$ are the (next-)nearest neighbors hopping
amplitudes. We choose $J_{j,j'}=J$ $(J_{j,j'}=J')$ for $j$ and
$j'$ within the same unit cell (in different unit cells), cf sketch
in Fig. \ref{fig:Optimization}(a). Likewise, we choose $L_{j,j'}=L$
$(L_{j,j'}=L')$ for $j$ and $j'$ within the same unit cell (in
different unit cells), cf sketch in Fig. \ref{fig:Optimization}(c). 

In order to analyse the (quasi-)BR as a function of the parameters
for our TB model, we need to calculate the spectrum and symmetry at
the high symmetry points. It is possible to find simple close formulas
because at each high symmetry point there are at most two orbitals
for each irreps. At $\Gamma$ point, the six sites combine to generate
one orbital for each quasi-angular momentum, the corresponding energies
are 
\begin{align*}
 & E_{s}/\hbar=\omega_{0}+3\bar{J}-\frac{\delta J}{2}-6\bar{L}-\delta L,\\
 & E_{p}/\hbar=\omega_{0}+\delta J+3\bar{L}+\frac{\delta L}{2},\\
 & E_{d}/\hbar=\omega_{0}-\delta J+3\bar{L}+\frac{\delta L}{2},\\
 & E_{f}/\hbar=\omega_{0}+3\bar{J}-\frac{\delta J}{2}-6\bar{L}-\delta L.
\end{align*}
At the $K$ points we have two orbitals for each values of the quasi-angular
momentum $m=0,\pm1$,
\begin{align}
E_{0/1,0}/\hbar & =\omega_{0}+2\delta L\mp\sqrt{3\bar{J}^{2}-3\bar{J}\delta J+\frac{7}{4}\delta J^{2}},\nonumber \\
E_{0/1,1}/\hbar & =\omega_{0}-\delta L\mp\sqrt{3\bar{J}^{2}+\frac{1}{4}\delta J^{2}}.\label{eq:EnergyK}
\end{align}
At the $M$ points where the proper group is ${\cal C}_{2\nu}$, it
is useful to label the orbitals according to the parity under the
two mirror symmetries. {[}We thus implicitly take into account that
the space group of the TB model is actually the Wallpaper group $p6m$
and not just $p6$.{]} We then find
\begin{align}
 & E_{+-}/\hbar=\omega_{0}-2\bar{J}-\bar{L}-\frac{3}{2}\delta L,\nonumber \\
 & E_{--}/\hbar=\omega_{0}+2\bar{J}-\bar{L}-\frac{3}{2}\delta L,\nonumber \\
 & E_{0/1,-+}/\hbar=\omega_{0}+\frac{\bar{J}}{2}+\frac{\bar{L}}{2}-\frac{\delta J}{4}+\frac{3}{4}\delta L\nonumber \\
 & \pm\left(\frac{9}{4}(\bar{J}-\bar{L})^{2}-\frac{5}{4}(\bar{J}-\bar{L})(\delta J-\delta L)+\frac{17}{16}(\delta J-\delta L)^{2}\right)^{1/2},\nonumber \\
 & E_{0/1,++}/\hbar=\omega_{0}-\frac{\bar{J}}{2}+\frac{\bar{L}}{2}+\frac{\delta J}{4}+\frac{3}{4}\delta L\nonumber \\
 & \pm\left(\frac{9}{4}(\bar{J}+\bar{L})^{2}-\frac{5}{4}(\bar{J}+\bar{L})(\delta J+\delta L)+\frac{17}{16}(\delta J+\delta L)^{2}\right)^{1/2}.\label{eq:EnergyM}
\end{align}
Here, we have defined 
\begin{align*}
 & \bar{J}=(J+J')/2,\quad\delta J=J'-J,\quad\bar{L}=(L+L')/2,\\
 & \delta L=L'-L.
\end{align*}
 While it would be interesting to use the above analytical expressions
to derive all possible topological and hybridization phases supported
by the our TB model, below, we focus on the parameter regime where
the TB model describe a small perturbation about the graphene TB model,
and, thus, $\bar{J}>0$ is the largest hopping amplitude, $\bar{J}\gg\bar{L},\delta J,\delta L$.
In this framework, it is not necessary anymore to distinguish between
states with different mirror symmetry at the $M$ point. Instead,
we adopt the same convention used in the main text of ordering the
three odd (even) states under ${\cal C}_{2}$ rotations by increasing
energy,
\begin{align}
E_{0,-}/\hbar & \equiv E_{+-}/\hbar=\omega_{0}-2\bar{J}-\bar{L}-\frac{3}{2}\delta L,\nonumber \\
E_{0,+}/\hbar & \equiv E_{0,++}/\hbar\approx\omega_{0}-2\bar{J}-\bar{L}+\frac{2}{3}\delta J-\frac{7}{6}\delta L,\nonumber \\
E_{1,-}/\hbar & \equiv E_{0,-+}/\hbar\approx\omega_{0}-\bar{J}+2\bar{L}+\frac{1}{6}\delta J+\frac{1}{3}\delta L,\nonumber \\
E_{1,+}/\hbar & \equiv E_{1,++}/\hbar\approx\omega_{0}+\bar{J}+2\bar{L}-\frac{1}{6}\delta J+\frac{1}{3}\delta L,\nonumber \\
E_{2,-}/\hbar & \equiv E_{1,-+}/\hbar\approx\omega_{0}+2\bar{J}-\bar{L}-\frac{2}{3}\delta J+\frac{7}{6}\delta L\nonumber \\
E_{2,+}/\hbar & \equiv E_{--}/\hbar=\omega_{0}+2\bar{J}-\bar{L}-\frac{3}{2}\delta L.\label{eq:EnergyM-1}
\end{align}
Here, we have expanded Eq. (\ref{eq:EnergyM}) up to leading order.
Likewise, expanding Eq. (\ref{eq:EnergyK}) we find
\begin{align}
E_{0/1,0}/\hbar & =\omega_{0}+2\delta L\mp\sqrt{3}\left(\bar{J}-\frac{\delta J}{2}\right),\nonumber \\
E_{0/1,1}/\hbar & =\omega_{0}-\delta L\mp\sqrt{3}\bar{J}.\label{eq:EnergyK-1}
\end{align}

We first analyze the special case $\delta J=\delta L=0$. In this
case the smallest possible unit cell contains only two sites and we
recover the band structure of graphene (with next nearest neighbor
hopping) but folded into a smaller Brillouin zone (because in real
space we are using a larger unit cell containing six atoms). The two
connected bands of graphene give rise to six connected bands after
folding. Due to this underlying symmetry we also expect two triply
degenerate levels at the $K$ point. The reason is that three quasi-momenta
of the larger BZ that are mapped onto each other by ${\cal C}_{3}$
rotations are projected onto the same quasi-momentum of the smaller
BZ. Indeed from Eq. (\ref{eq:EnergyK}) we see that $E_{0,0}=E_{0,1}$
($E_{1,0}=E_{1,1}$) corresponding to a triple degeneracy because
$E_{0,1}$ ($E_{1,1}$) is a doubly degenerate level. With similar
arguments, one can prove that at the $\text{M}$ point two doubly
degenerate levels are to be expected. Indeed, from Eq. (\ref{eq:EnergyM})
we see that $E_{0,-}=E_{0,+}$ and $E_{2,-}=E_{2,+}$. 

\begin{figure}
\includegraphics[width=1\columnwidth]{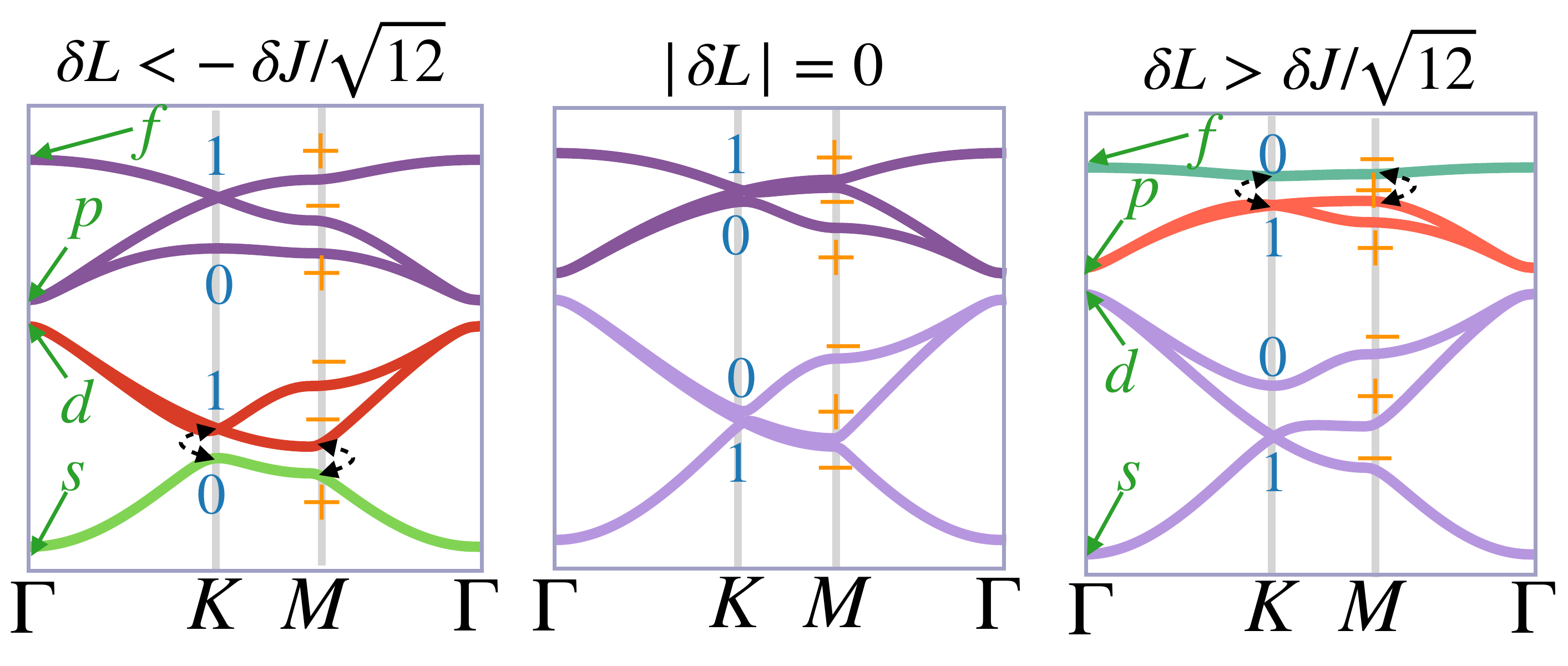}

\caption{\label{fig:fragile_TB_model}Three different phases for the graphene
TB model with next-nearest neighbor hopping. For $\delta J>0,$ $|\delta L|=0$
the folded graphene band structure is split into two Kagome like band
structures (central panel). For $\delta L<-\delta J/\sqrt{12}$, after
band inversion transitions at both the $M$ and the $K$ points (the
orbitals that are exchanged are marked by dashed lines), the lower
Kagome bands split into a triangular like $s$-band and a pair of
topological bands (left panel). For $\delta L>\delta J/\sqrt{12}$,
similar band inversions lead to the splitting of the higher Kagome
bands into a topological pair of bands and triangular like $f$-band
(right panel).}
\end{figure}

The first step towards constructing the topological quasi BRs is to
create an imbalance in the nearest neighbor hopping by choosing $\delta J>0$
(the external hopping is larger). In this scenario (discussed also
above for $\bar{L}=0$), the folded graphene band structure is split
into two sets of three connected bands each with dimerized Wannier
orbitals localized about the ${\cal C}_{2}$ roto-centers, cf Fig.
\ref{fig:fragile_TB_model} (central panel). At the $\Gamma$ point
the $p$ orbital is lifted above the $d$ orbital. At the same time
at the $M$-point, the lowest (highest) energy even band is lifted
above the lowest (highest) energy odd band, $E_{0,+}>E_{0,-}$ ($E_{2,+}>E_{2,-}$
). Likewise, at the $K$ point the lowest (highest) $s$-orbital wave
is lifted above (lowered below) the lowest (highest) $p$-level, $E_{0,0}>E_{0,1}$
($E_{1,0}<E_{1,1}$). 

Next we tweak the band structure described above to obtain topological
fragile bands. This is achieved by creating an imbalance between the
next-nearest neighbor hopping, $\delta J\neq0$. From Eq. (\ref{eq:EnergyM-1})
we see that at the $M$-point, a positive $\delta J$ decreases the
energy $E_{2,+}$ of the highest even orbital while increasing the
energy of the odd orbital $E_{2,-}$. Meanwhile at $K$-point, the
energy $E_{1,1}$ of the highest $p$-Bloch wave is also decreased
while the energy $E_{1,0}$ of the corresponding $s$-orbital is increased,
cf Eq. (\ref{eq:EnergyK-1}). For sufficiently large $\delta L$,
$\delta L>\delta J/\sqrt{12}$, the order of the highest two bands
have been inverted compared to the situation where $\delta L=0$ at
both high symmetry points $K$ and $M$ (at the $M$ point the band
inversion occurs already for $\delta L>\delta J/4$). As a consequence,
the highest three bands are split into into a pair of topological
bands and an $f$-orbital localized about the ${\cal C}_{6}$ rotocenter,
cf Fig. \ref{fig:fragile_TB_model} (right panel) and Fig. 4(c) of
the main text.

A similar analysis shows that for $\delta L$ negative, $\delta L<-\delta J/\sqrt{12}$,
the lowest three bands are split into a $s$-orbital localized about
the ${\cal C}_{6}$ rotocenters and a pair of topological bands, cf
Fig. \ref{fig:fragile_TB_model} (left panel). This is similar to
what is observed for the lowest three bands of the randomly generated
potential 2 in Fig. 3 of the main text. 

\bibliographystyle{apsrev4-1}
\bibliography{NNBandstructures}

\end{document}